%% file: main.tex
\documentclass[twocolumn]{autart} 
\input{comande.tex}

\usepackage{ mathrsfs }
\usepackage{ upgreek }
\usepackage{subcaption}
\newtheorem{stan_assum}{Standing Assumption}

\begin{document}

\begin{frontmatter}

\title{Learning generalized Nash equilibria in monotone games: A hybrid adaptive extremum seeking control approach} 

\thanks[footnoteinfo]{This work was partially supported by the ERC under research project COSMOS (802348). E-mail addresses: \texttt{\{s.krilasevic-1, s.grammatico\}@tudelft.nl.}}

\author{Suad Krilašević} and
\author{\ Sergio Grammatico} 
          
\address{Delft Center for Systems and Control, TU Delft, The Netherlands}


\begin{abstract}                          
In this paper, we solve the problem of learning a generalized Nash equilibrium (GNE) in merely monotone games. First, we propose a novel continuous semi-decentralized solution algorithm without projections that uses first-order information to compute a GNE with a central coordinator. As the second main contribution, we design a gain adaptation scheme for the previous algorithm in order to alleviate the problem of improper scaling of the cost functions versus the constraints. Third, we propose a data-driven variant of the former algorithm, where each agent estimates their individual pseudogradient via zeroth-order information, namely, measurements of their individual cost function values. Finally, we apply our method to a perturbation amplitude optimization problem in oil extraction engineering. 
\end{abstract}

\end{frontmatter}

\section{Introduction}
Decision problems where self-interested intelligent systems or agents wish to optimize their individual cost objective function arise in many engineering applications, such as charging/discharging coordination for plug-in electric vehicles \cite{ma2011decentralized}, \cite{grammatico2017dynamic}, demand-side management in smart grids \cite{mohsenian2010autonomous}, \cite{saad2012game}, robotic formation control \cite{lin2014distributed} and thermostatically controlled loads \cite{li2015market}. The key feature that distinguishes these problems from multi-agent distributed optimization is the fact the cost functions and constraints are coupled together. Currently, one active research area is that of finding (seeking) actions that are self-enforceable, e.g. actions such that no agent has an incentive to unilaterally deviate from - the so-called generalized Nash equilibrium (GNE) \cite[Eq. 1]{facchinei2010generalized}. Due to the aforementioned coupling, information on other agents must be communicated, observed, or measured in order to compute a GNE algorithmically.  The nature of this information can vary from knowing everything (full knowledge of the agent actions) \cite{yi2019operator}, estimates based on distributed consensus between the agents \cite{gadjov2019distributed}, to payoff-based estimates \cite{marden2009cooperative}, \cite{frihauf2011nash}. The latter is of special interest as it requires no dedicated inter-agent communication infrastructure.  \\ \\

\emph{Literature review:} In payoff-based algorithms, each agent can only measure the value of their cost function, but does not necessarily know its analytic form. Many of such algorithms are designed for Nash equilibrium problems (NEPs) with finite action spaces where each agent has a fixed policy that specifies what a player should do under any condition, e.g. \cite{goto2012payoff}, \cite{marden2009cooperative}, \cite{marden2012revisiting}. On the other hand, the main component of continuous action space algorithms is the payoff-based (pseudo)gradient estimation scheme. A notable class of payoff-based algorithms called Extremum Seeking Control (ESC) is based on the seminal work by Krstić and Wang  \cite{krstic2000stability}. The main idea is to use perturbation signals to ``excite'' the cost function and estimate its gradient which is then used in a gradient-descent-like algorithm. Since then, various different variants have been proposed \cite{liu2011stochastic}, \cite{ghaffari2012multivariable}, \cite{durr2013lie}, \cite{grushkovskaya2018class}, \cite{liao2019constrained}, \cite{shao2019extremum}, \cite{labar2019newton}.  A full-information algorithm where the (pseudo)gradient is known, can be ``transformed" into an extremum seeking one if it satisfies some properties, like continuity of the dynamics, use of only one (pseudo)gradient in the dynamics, appropriate stability of the optimizer/NE, etc. At first, (local) exponential stability of the optimizer/NE was assumed or implied with other assumptions \cite[Assum. 2.2]{krstic2000stability}, \cite[Assum. 3.1]{frihauf2011nash}. Thanks to results in averaging and singular perturbation theory \cite{sanfelice2011singular},\cite{wang2012analysis} in the hybrid dynamical systems framework \cite{goebel2012hybrid}, the assumption was relaxed to just (practical) asymptotic stability \cite{poveda2017framework}. Subsequently, extremum seeking algorithms were developed for many different applications, such as event-triggered optimization \cite{poveda2017robust}, Nesterov-like accelerated optimization with resetting \cite{poveda2021robust}, optimization of hybrid plants \cite{poveda2018hybrid}, population games \cite{poveda2015shahshahani}, N-cluster Nash games \cite{ye2020extremum}, fixed-time Nash equilibrium seeking for strongly monotone games \cite{poveda2020fixed}, Nash equilibrium seeking for merely monotone games \cite{krilavsevic2021extremum} and generalized Nash equilibrium seeking in strongly monotone games \cite{krilavsevic2021learning}.\\ \\

GNEPs can be solved efficiently by casting them into a variational inequality (VI) \cite[Equ. 1.4.7]{facchinei2007finite}, and in turn into the problem of finding a zero of an operator \cite[Equ. 1.1.3]{facchinei2007finite}, for which there exists a vast literature \cite{bauschke2011convex}. For GNEPs, this operator is the KKT operator, composed of the pseudogradient (whose monotonicity determines the type of the game), dual variables, constraints and their gradients. In the case of merely monotone operators, the most widely used solution algorithms are the forward-backward-forward \cite[Rem. 26.18]{bauschke2011convex}, the extragradient \cite{korpelevich1976extragradient} and the subgradient extragradient \cite{censor2011subgradient}. The main drawback of all of these algorithms, with respect to an extremum seeking adaptation, is that they require two pseudogradient computations per iteration. Recently, the golden ratio algorithm has been proven to converge in the monotone case with only one pseudogradient computation \cite{malitsky2019golden}. There also exist continuous-time versions of the aforementioned algorithms, like the forward-backward-forward algorithm \cite{bot2020forward} and the golden ratio algorithm \cite{gadjov2020exact}, albeit without projections in the latter case, rendering it unusable for GNEPs, as projections are essential for the dual dynamics. To the best of our knowledge, in the merely monotone case, there currently exist no continuous-time GNEP algorithm that can be paired with extremum seeking. \\

\emph{Contribution}: Motivated  by  the  above  literature  and open  research  problem,  to  the  best  of  our  knowledge, we consider and solve the problem of learning (i.e., seeking via zeroth-order information) a GNE in merely monotone games. Specifically, our main technical contributions are summarized next:

\begin{itemize}
    \item We propose a novel, semi-decentralized, single pseudogradient computation generalized Nash equilibrium (GNE)  seeking algorithm for solving monotone GNEPs, inspired by the golden ratio algorithm in \cite{malitsky2019golden}, \cite{gadjov2020exact}.
    \item We propose a novel dual variable gain adaptation scheme using the framework of hybrid dynamical systems in order to alleviate the problem of improper scaling of the cost and constraint functions.  
    \item We propose a novel extremum seeking scheme which exploits the aforementioned properties of the previous algorithms and applies it to GNEPs. 
\end{itemize}
\emph{Comparison with \cite{krilavsevic2021extremum} and \cite{krilavsevic2021learning}:} We emphasize that since here we assume non-strong monotonicity of the pseudogradient mapping, the methodology in  \cite{krilavsevic2021learning} based on the forward-backward splitting is not applicable - see \cite[Equ. 4]{gadjov2020exact} for an example of non-convergence. Furthermore, by incorporating projectionless dual dynamics, here we allow for the existence of constraints, in contrast with the methodology in \cite{krilavsevic2021extremum} which cannot be extended to the constrained case. Thus, in this paper, we develop a novel splitting methodology that solves the issues of non-convergence and constrained feasible set, and consequently addresses a much wider class of equilibrium problems. Moreover, the hybrid gain adaptation is also novel and not considered in these previous works. \\
\emph{Notation}: $\mathbb{R}$ denotes the set of real numbers. For a matrix $A \in \mathbb{R}^{n \times m}$, $A^\top$ denotes its transpose. For vectors $x, y \in \mathbb{R}^{n}$ and $M \in \R^{n \times n}$ a positive semi-definite matrix and $\mathcal{A} \subset \R^n$, $\vprod{x}{y}$, $\|x \|$,  $\|x \|_M$ and $\|x \|_\cal{A}$ denote the Euclidean inner product, norm, weighted norm and distance to set respectively. Given $N$ vectors $x_1, \dots, x_N$, possibly of different dimensions, $\col{x_1, \dots x_N} \coloneqq \left[ x_1^\top, \dots, x_N^\top \right]^\top $. Collective vectors are defined as $\bfs{x} \coloneqq \col{x_1, \dots, x_N}$ and for each $i = 1, \dots, N$, $\bfs{x}_{-i} \coloneqq \col{ x_1, \dots,  x_{i -1},  x_{i + 1}, \dots, x_N } $. Given $N$ matrices $A_1$, $A_2$, \dots, $A_N$, $\operatorname{diag}\left(A_{1}, \ldots, A_{N}\right)$ denotes the block diagonal matrix with $A_i$ on its diagonal. For a function $v: \mathbb{R}^{n} \times \mathbb{R}^{m}  \rightarrow \mathbb{R}$ differentiable in the first argument, we denote the partial gradient vector as $\nabla_x v(x, y) \coloneqq \col{\frac{\partial v(x, y)}{\partial x_{1}}, \ldots, \frac{\partial v(x, y)}{\partial x_{N}}} \in \mathbb{R}^{n}$. We use $\mathbb{S}^{1}:=\left\{z \in \mathbb{R}^{2}: z_{1}^{2}+z_{2}^{2}=1\right\}$ to denote the unit circle in $\R^2$. 
$\operatorname{Id}$ is the identity operator. $I_n$ is the identity matrix of dimension $n$ and $ \bfs{0}_n$ is vector column of $n$ zeros. A continuous function $\gamma: \R_+ \leftarrow \R_+$ is of class $\mathcal{K}$ if it is zero at zero and strictly increasing. A continuous function $\alpha: \R_+ \leftarrow \R_+$ is of class $\mathcal{L}$ if is non-increasing and converges to zero as its arguments grows unbounded. A continuous function $\beta: \R_+ \times \R_+ \rightarrow \R_+$ is of class $\mathcal{KL}$ if it is of class $\mathcal{K}$ in the first argument and of class $\mathcal{L}$ in the second argument.\\ \\

The framework of hybrid dynamical systems (HDS) theory \cite{goebel2012hybrid} like \cite{sanfelice2011singular}, \cite{wang2012analysis}, \cite[Lemma 4]{poveda2017framework} is especially attractive for extremum seeking, as it allows one to quickly and elegantly prove various stability theorems \cite{poveda2017framework}, \cite{poveda2017robust}, \cite{poveda2020fixed}, \cite{poveda2021robust}. Thus, we also use the framework of HDSs to model our algorithms. A HDS is defined as \\
\centerline{
\begin{minipage}[c]{0.3\textwidth}
\begin{subequations}
\begin{align}
    &\dot{x} \in F(x) & \text{if}\quad &x \in C \label{eq: flow map def}\\ 
    &x^+ \in G(x) & \text{if}\quad &x \in D, \label{eq: jump map def}
\end{align}\label{eq: HDS}
\end{subequations}
\end{minipage}} \\


where $x \in \R^{n}$ is the state, $F: \R^{n} \rightarrow \R^{n}$ is the flow map, and $G: \R^{n} \rightarrow \mathbb{R}^{n}$ is the jump map, the sets $C$ and $D$, are the flow set and the jump set, respectively, that characterize the points in space where the system evolves according to \nref{eq: flow map def}, or \nref{eq: jump map def}, respectively. The data of the HDS is defined as $\mathcal{H}:=\{C, D, F, G\}$. Solutions $x:\dom(x) \rightarrow \R^{n}$ to \nref{eq: HDS} are defined on hybrid time domains, and they are parameterized by a continuous-time index $t \in \R_+$ and a discrete-time index $j \in \mathbb{Z}_+$. Solutions with unbounded time or index domains are said to be complete \cite[Chp. 2]{goebel2012hybrid}. We now define various forms of stability and other basic concepts for HDSs.

\begin{defn}[UG(p)AS] A compact set $\mathcal{A} \subset \mathbb{R}^{n}$ is said to be Uniformly Globally pre-Asymptotically Stable (UGpAS) for a HDS $\mathcal{H}$ if there exists $\beta \in \mathcal{K} \mathcal{L}$ such that every solution $x$ of $\mathcal{H}$ satisfies $\n{x(t, j)}_{\mathcal{A}} \leq \beta\left(\n{x(0,0)}_{\mathcal{A}}, t+j\right)$, for all $(t, j) \in \operatorname{dom}(x)$. If additionally all solutions are complete, we then use the acronym UGAS. \kraj
\end{defn}

\begin{defn}[SG(p)AS] For a parameterized HDS $\mathcal{H}_{\varepsilon}$, $\varepsilon \in \R^k_+$, a compact set $\mathcal{A} \subset \mathbb{R}^{n}$ is said to be Semi-Globally Practically pre-Asymptotically Stable (SGPpAS) as 
$(\varepsilon_1, \dots, \varepsilon_k) \rightarrow 0^+$ with $\beta \in \mathcal{K} \mathcal{L}$ if for all compact sets $K \subset \mathbb{R}^{n}$ and all $v>0$, there exists $\varepsilon_{0}^{*}>0$ such that for each $\varepsilon_{0} \in\left(0, \varepsilon_{0}^{*}\right)$ there exists $\varepsilon_{1}^{*}\left(\varepsilon_{0}\right)>0$ such that for each $\varepsilon_{1} \in$
$\left(0, \varepsilon_{1}^{*}\left(\varepsilon_{0}\right)\right) \ldots$ there exists $\varepsilon_{j}^{*}\left(\varepsilon_{j-1}\right)>0$ such that for each $\varepsilon_{j} \in$
$\left(0, \varepsilon_{j}^{*}\left(\varepsilon_{j-1}\right)\right) \ldots, j=\{2, \ldots, k\},$  every solution $x_{\varepsilon}$ of $\mathcal{H}_{\varepsilon}$
with $x_{\varepsilon}(0,0) \in K$ satisfies
$$\n{x_{\varepsilon}(t, j)}_{\mathcal{A}} \leq \beta\left(\n{x_{\varepsilon}(0,0)}_{\mathcal{A}}, t+j\right)+v$$
for all $(t, j) \in \operatorname{dom}\left(x_{\varepsilon}\right)$. If additionally all solutions are complete, we then use the acronym SGPAS. \kraj
\end{defn}

\begin{defn}[Hybrid basic conditions]
A HDS in \nref{eq: HDS} is said to satisfy the Hybrid basic conditions if $C$ and $D$ are closed, $C \subset \dom(F)$, $D \subset \dom(G)$, $F$ and $G$ are continuous on $C$ and $D$ respectively. \kraj
\end{defn}

Furthermore, let us define a notion of robustness with respect to small disturbances as in \cite{poveda2021robust}:

\begin{defn}[Structural robustness]Let $\mathcal{H}$ render UGpAS (resp. SGPpAS as $\varepsilon \rightarrow 0^{+}$ ) a compact set $\mathcal{A}$ with $\beta \in \mathcal{K} \mathcal{L}$. We say that $\mathcal{H}$ is Structurally Robust if for all measurable functions $e: \mathbb{R}_{\geq 0} \rightarrow \mathbb{R}^{n}$ satisfying $\sup _{t \geq 0}\n{e(t)} \leq \bar{e}$, with $\bar{e}>0$, the perturbed system

\begin{subequations}
\begin{align}
    &x+e \in C, &\dot{x}=F(x+e)+e \\
    &x+e \in D, &x^{+}=G(x+e)+e
\end{align}
\end{subequations}

renders the set $\mathcal{A}$ SGPpAS as $\bar{e} \rightarrow 0^{+}$ (resp. SGPpAS as $(\varepsilon, \bar{e}) \rightarrow$ $\left.0^{+}\right)$ with $\beta \in \mathcal{K} \mathcal{L}$. \kraj
\end{defn}

\section{Generalized Nash equilibrium problem}

We consider a multi-agent system with $N$ agents indexed by $i \in \mathcal{I} \coloneqq \{1, 2, \dots N\}$, each with cost function 
\begin{align}
    J_i(u_i, \boldsymbol{u}_{-i}), \label{eq: cost functions}
\end{align}

where $u_i \in \R^{m_i}$ is the decision variable,  $J_i: \R_{m_i} \times \R_{m_{-i}} \rightarrow \R$. Let us also define $m \coloneqq \sum_{j \in \mathcal{I}} m_j$ and $m_{-i} \coloneqq \sum_{j \neq i} m_j$. Formally, we do not consider local constraints, but they can be incorporated softly into the cost function via penalty-barrier functions. All agents are subject to convex coupling constraints $g_j(\bfs{u})$ indexed by $j \in \mathcal{Q} \coloneqq \{1, 2, \dots q\}$. Therefore, let us denote the overall feasible decision set as 
\begin{align}
    &\bfs{\mathcal{U}}\coloneqq\left\{\bfs{u} \in \mathbb{R}^{m}\ |\  g(\bfs{u}) \leq \bfs{0} \right\}, \label{def: U}
\end{align}{}%
and the feasible set of agent $i$ as
\begin{align}
    &{\mathcal{U}}_i(\bfs{u}_{-i}) \coloneqq \left\{{u}_i \in \mathbb{R}^{m_i}\ |\ g(\bfs{u}) \leq \bfs{0} \right\},
\end{align}
where $g(\bfs{u}) = \col{\z{g_j(\bfs{u})}_{j \in \mathcal{Q}}}$.\\

The goal of each agent is to minimize their cost function, i.e.,
\begin{align}
\forall i \in \mathcal{I}:\ &\min_{u_i \in \mathcal{U}_i(\bfs{u}_{-i})}  J_i(u_i, \boldsymbol{u}_{-i}), \label{def: game}
\end{align}{}
which depends on the decision variables of other agents as well. Thus, a game $\mathcal{G}$ is defined by the set of cost functions and the feasible set, i.e. $\mathcal{G} \coloneqq \{(J_i(\bfs{u}))_{i \in \mathcal{I}}, (g_j(\bfs{u}))_{j \in \mathcal{Q}}\}$. From a game-theoretic perspective, this is the problem to compute a generalized Nash equilibrium (GNE), as formalized next. \\
\begin{defn}[Generalized Nash equilibrium]\hfill\quad
A set of control actions $\bfs{u}^*\coloneqq\col{u_i^*}_{i \in \mathcal{I}}$ is a generalized Nash equilibrium if, for all $i \in \mathcal{I}$,
\begin{align}
    u_{i}^{*} \in \underset{v_{i}}{\operatorname{argmin}}\ J_{i}\left(v_{i}, \bfs{u}_{-i}^{*}\right)\, \mathrm{ s.t. }\left(v_{i}, \bfs{u}_{-i}^{*}\right) \in \bfs{\mathcal{U}}.  \label{def: gne}
\end{align}
with $J_i$ as in \nref{eq: cost functions} and $\bfs{\mathcal{U}}$ as in \nref{def: U}.\kraj \\ \\
\end{defn}{}
In plain words, a set of inputs is a GNE if no agent can improve their cost function by unilaterally changing their input. \\ \\

A common approach for solving a GNEP is to translate it into a quasi-variational inequality (QVI) \cite[Thm. 3.3]{facchinei2010generalized} that can be simplified to a variational inequality (VI) \cite[Thm. 3.9]{facchinei2010generalized} for a certain subset of solutions called variational-GNE (v-GNE), which in turn can be translated into a problem of finding zeros of a monotone operator \cite[Equ. 1.1.3]{facchinei2007finite}. To ensure the equivalence of the GNEP and QVI, we postulate the following assumption \cite[Thm. 3.3]{facchinei2010generalized}:
\begin{stan_assum}[Regularity] \label{sassum: regularity}
For each $i \in \mathcal{I}$, the function $J_i$ in \nref{eq: cost functions} is continuous; the function $J_{i}\left(\cdot, \bfs{u}_{-i}\right)$ is convex for every $\bfs{u}_{-i}$; For each $j \in \mathcal{Q}$, convex constraint $g_j(\bfs{u})$ is continuously differentiable, $\bfs{\mathcal{U}}$ is non-empty and satisfies Slater's constraint qualification. \kraj
\end{stan_assum}{}
We focus on a subclass of GNE called variational GNE \cite[Def. 3.10]{facchinei2010generalized}. A collective decision $\bfs{u}^*$ is a v-GNE in \nref{def: gne} if and only if there exists a dual variable $\lambda^* \in \R^q$ such that the following KKT conditions are satisfied \cite[Th. 4.8]{facchinei2010generalized}:

\begin{align}
    \mathbf{0}_{m + q} &\in F_{\text{ex}}(\bfs{u}^*, \lambda^*) \coloneqq \m{F\left(\bfs{u}^*\right)+\nabla g(\bfs{u}^*)^{\top} \lambda^* \\ -g(\bfs{u}^*) + \mathrm{N}_{\R^q_+}(\lambda^*)}, \label{eq: kkt} 
\end{align}
where by stacking the partial gradients $\nabla_{u_i} J_i(u_i, \boldsymbol{u}_{-i})$ into a single vector, we have the so-called pseudogradient mapping:
\begin{align}
    F(\boldsymbol{u}):=\operatorname{col}\left(\left(\nabla_{u_{i}} J_{i}\left(u_{i}, \bfs{u}_{-i}\right)\right)_{i \in \mathcal{I}}\right). \label{eq: pseudogradient}
\end{align}
Let us also postulate the weakest working assumption in GNEPs with continuous actions, i.e. the monotonicity of the pseudogradient mapping  \cite[Def. 2.3.1, Thm. 2.3.4]{facchinei2007finite}.
\begin{stan_assum}[Monotonicity]
The pseudogradient mapping $F$ in \nref{eq: pseudogradient} is monotone, i.e., for any pair\, $\bfs{u}, \bfs{v} \in \bfs{\Omega}$, it holds that $\vprod{\bfs{u} - \bfs{v}}{F(\bfs{u}) - F(\bfs{v})} \geq 0$. \kraj
\end{stan_assum}{}
The regularity and monotonicity assumptions are not enough to ensure the existence of a v-GNE \cite[Thm. 2.3.3, Corr. 2.2.5]{facchinei2007finite}, \cite[Thm. 6]{facchinei2010generalized}, hence let us postulate its existence:
\begin{stan_assum}[Existence] \label{sassum: exist}
There exists $\omega^* \coloneqq \col{\bfs{u}^*, \lambda^*} \in \R^m \times \R_+^q$ such that Equation \nref{eq: kkt} is satisfied.\kraj
\end{stan_assum}{}

In this paper, we consider the problem of finding a GNE of the game in \nref{def: game} via zeroth-order information, i.e. local measurements of the cost functions in \nref{eq: cost functions}.
\section{Full-information generalized Nash equilibrium seeking}
We present two novel full-information GNE seeking algorithms. In the first algorithm, the dual variables are calculated without the use of projections by a central coordinator. The lack of projections onto tangent cones, along with the fact that the flow map of the algorithm contains only one pseudogradient computation and that the algorithm itself converges merely under the monotonicity assumption, enables us to use hybrid dynamical system theory for the zeroth-order extension of the algorithm later on. In the second algorithm, we propose a hybrid gain adaptation scheme, in order to improve the performance of the algorithm when we do not know a priori how to best tune the gains.

\subsection{Projectionless GNE seeking algorithm}\label{sec: projectionless gne}

The algorithm in \cite{gadjov2020exact} proves convergence to a NE for a monotone pseudogradient by combining an additional filtering dynamics with the standard NE seeking one. Similarly, we propose a Lagrangian first-order primal dynamics with filtering for each agent:
\begin{align}
    \m{ \dot{u}_i \\ \dot{z}_i } = \m{- u_i + z_i - \gamma_i\z{\nabla_{u_i}J_i(u_i, \bfs{u}_{-i}) + \nabla_{u_i} g(\bfs{u})^\top \lambda } \\ - z_i + u_i }. \nonumber
\end{align}{}
The authors in \cite{gadjov2020exact} propose a passivity framework for the convergence of their algorithm. Instead, we offer a different intuition for the convergence. The additional dynamics make impossible any $\omega$-limit trajectories other than that of stationary points for which the flow map is equal to zero, i.e., there cannot be any ``movement" in the invariant set, thus enabling convergence under merely the monotonicity assumption. In the case of the dual dynamics, in order to avoid projections, we propose the following dynamics:
\begin{align}
    \forall j \in \mathcal{Q}: \dot\lambda_j &= \lambda_j\z{g_j(\bfs{u}) - \lambda_j + w_j} \red
    \dot w &= - w + \lambda. \label{eq: dual dynamics}
\end{align}
While the classic dual Lagrangian dynamics preserve the positivity of the dual variables by projecting onto the positive orthant, the same is accomplished in \nref{eq: dual dynamics} by ``slowing down" the dynamics of each individual dual variable proportionally to their distance to zero. Unlike \cite{durr2011smooth}, \cite{durr2013saddle}, where strict convexity of the cost and constraint functions is assumed to avoid the problem with $\omega$-limit trajectories in the invariant set, thanks to our newfound understanding of the filtering dynamics, we incorporate it to eliminate the strict convexity assumption. \\ \\
Thus, in collective form, we have
\begin{align}
    \dot{\omega} = \m{ \dot{\bfs{u}} \\ \dot{\bfs{z}} \\ \dot\lambda  \\ \dot{{w}} } = \m{-\bfs{u} + \bfs{z} - \Gamma (F(\bfs{u}) + \nabla g(\bfs{u})^\top \lambda )\\ -\bfs{z} + \bfs{u}\\ \diag{\lambda} \z{g(\bfs{u}) - \lambda + w} \\ -{w} + \lambda }\label{eq: full info dynamics}
\end{align}

Let us define the set of equilibrium points of the dynamics in \nref{eq: full info dynamics} as
\begin{align}
    &\mathcal{M} \coloneqq \Big\{\omega \in \R^{2m} \times \R_+^{2q} \mid \bfs{u} = \bfs{z}, w = \lambda, \mathbf{0}_{m} = F\left(\bfs{u}\right)\red 
    &+\nabla g(\bfs{u})^{\top} \lambda, \diag{\lambda} \diag{g(\bfs{u})} = 0\Big\},
\end{align}
its subset $\mathcal{A}$ which relates to the solutions of the game in \nref{def: game} as
\begin{align}
    &\mathcal{A} \coloneqq \Big\{\omega \in \R^{2m} \times \R_+^{2q} \mid \bfs{u} = \bfs{z}, w = \lambda, \red 
    &\mathbf{0}_{m} \in  F_{\text{ex}}\left(\bfs{u}, \lambda\right)\Big\}\subseteq \cal{M} \label{eq: equilibrium points},
\end{align}

and $\mathcal{L}$ as the set where at least one dual variable is equal to zero:
\begin{align*}
    \mathcal{L} \coloneqq \{\omega \in \R^{2m} \times \R_+^{2q} \mid \lambda_1 \cdot \lambda_2 \cdot \dotsc \cdot \lambda_q = 0 \}.
\end{align*}

\begin{figure*}
\centering
        \begin{subfigure}[t]{0.18\textwidth}
            \centering
            \includegraphics[width=\textwidth]{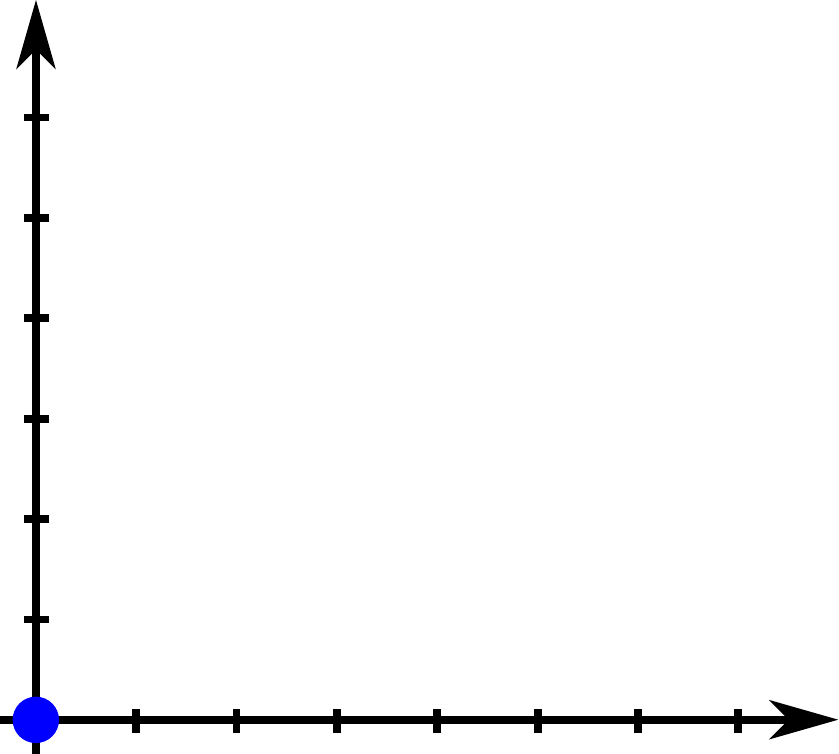}
            \caption{$\mathcal{A}$ (blue dot) contains a single point.}    
            \label{fig:set_noconstraints}
        \end{subfigure}
        \hspace{1em}
        \begin{subfigure}[t]{0.18\textwidth}   
            \centering 
            \includegraphics[width=\textwidth]{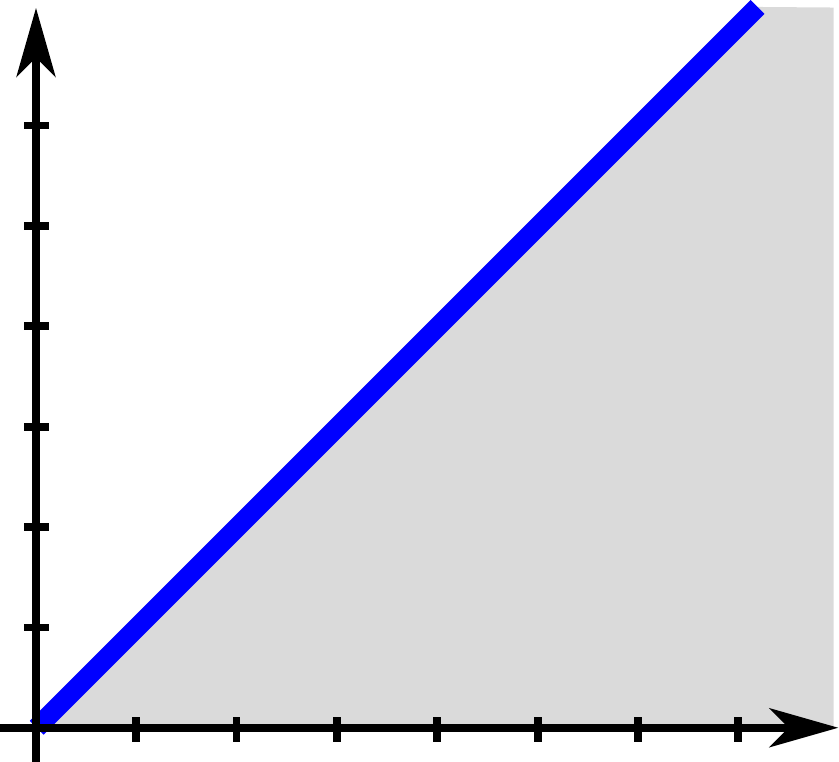}
            \caption{$\mathcal{A}$ (blue line) contains multiple points.}  
            \label{fig:set_many}
        \end{subfigure}
        \hspace{1em}
        \begin{subfigure}[t]{0.18\textwidth}
            \centering
            \includegraphics[width=\textwidth]{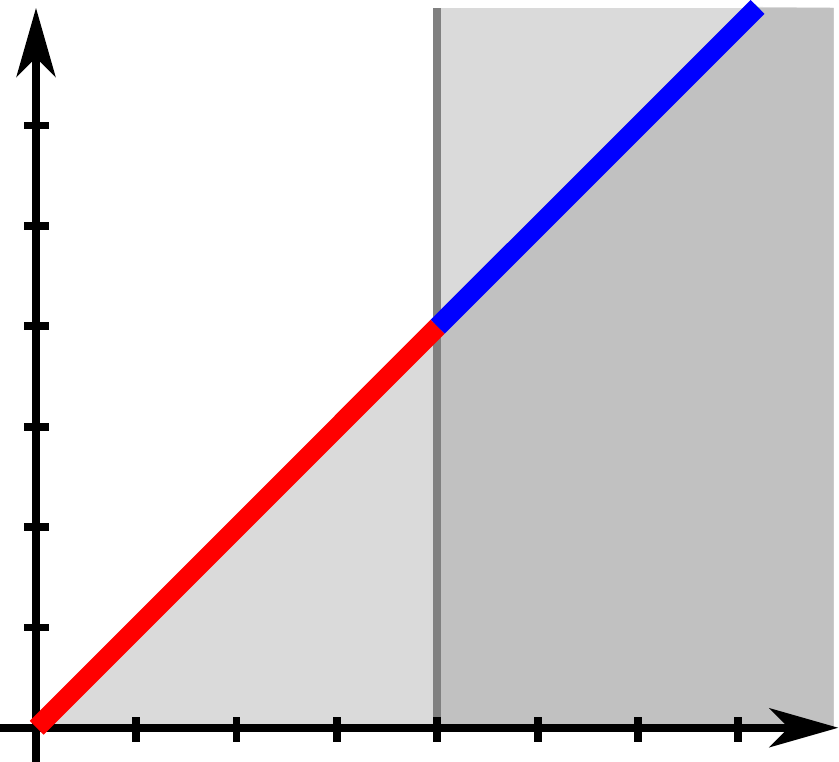}
            \caption{$\mathcal{A}$ (blue line) and $\mathcal{M} \setminus \mathcal{A}$ (red line) are connected.}    
            \label{fig:set_connected}
        \end{subfigure}
        \hspace{1em}
        \begin{subfigure}[t]{0.18\textwidth}   
            \centering 
            \includegraphics[width=\textwidth]{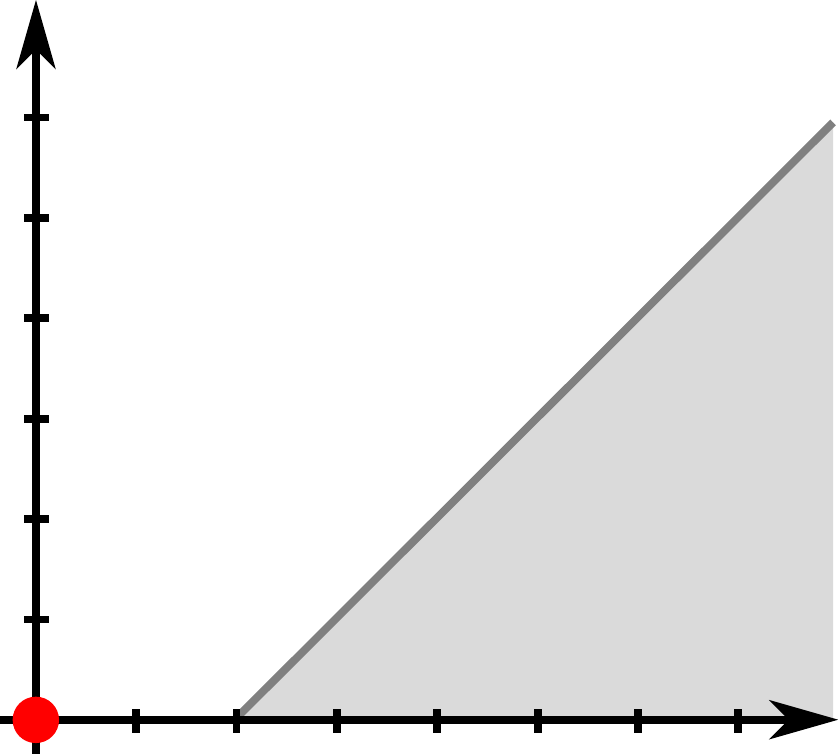}
            \caption{$\mathcal{A}$ contains no points, set $\cal{M}$ (red dot) is a single point.}  
            \label{fig:set_none}
        \end{subfigure}
        \hspace{1em}
        \begin{subfigure}[t]{0.18\textwidth}  
            \centering 
            \includegraphics[width=\textwidth]{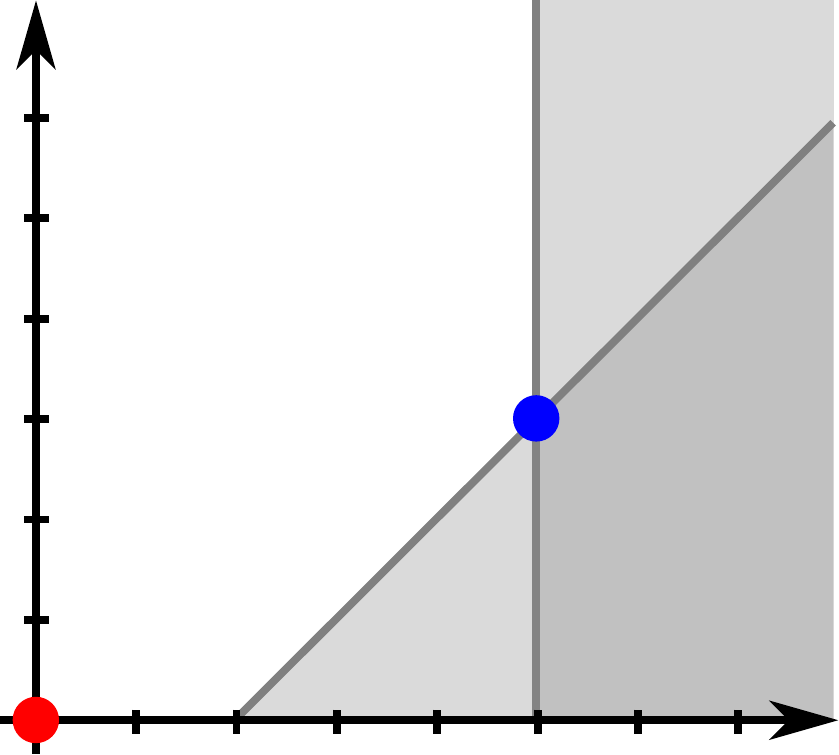}
            \caption{$\mathcal{A}$ (blue dot) and $\mathcal{M} \setminus \mathcal{A}$ (red dot) disconnected.}  
            \label{fig:set_disconnected}
        \end{subfigure}
        
        \caption{Solutions to several game scenarios with $F(\bfs{u}) \coloneqq \col{u_2, -u_1}$: $\mathcal{A}$ is shown in blue, while the other equilibrium points of \nref{eq: full info dynamics} $\mathcal{M}\setminus\mathcal{A}$,  are shown in red. Areas that satisfy the constraints are shown in gray.} 
        \label{fig:sets}
    \end{figure*}

Some example sets can be seen on Figure \ref{fig:sets}. As shown in Figure \ref{fig:set_disconnected}, the set $\mathcal{M}$ is not necessarily connected. Without constraints, $\cal{M} $ is equivalent to $\cal{A}$ and it contains only the zeros of the pseudogradient as shown in Figure \ref{fig:set_noconstraints}. By adding constraints, we can either create new equilibrium solutions (Figures \ref{fig:set_many}, \ref{fig:set_disconnected}) or ``remove" previous ones (Figure \ref{fig:set_none}). Either way, ``all" the solutions are still included in the set $\cal{M}$, which is the union of all solutions to games $\{(J_i(\bfs{u}))_{i \in \mathcal{I}}, (g_j(\bfs{u}))_{j \in \tilde{\mathcal{Q}}}\}$, where $\tilde{\mathcal{Q}}$ is a subset of $\mathcal{Q}$. \\ \\
We later show that $\cal{M}$ is attractive. Additionally, the following Lemma characterizes the stability of points in $\cal{M}\setminus\mathcal{A}$.
\begin{lem}\label{lemma: unstability}
Let the Standing Assumptions hold. Then, the equilibrium points in $\mathcal{M}\setminus\mathcal{A}$ are unstable for dynamics in \nref{eq: full info dynamics}.
\end{lem}
\begin{pf}
See Appendix \ref{app: proof of lemma 1}.\krajdokaz
\end{pf}
Therefore, in order to prove stability of $\mathcal{A}$, we need the sets $\cal{A}$ and $\cal{M}\setminus\cal{A}$ to be disjoint. In Figures \ref{fig:set_many} and \ref{fig:set_connected} we illustrate this situation happens when the solution set contains multiple points and some of them are ``removed" by the introduction of the new constraints. Thus, we have to assume this is not the case:

\begin{stan_assum}[Isolation of solutions] \label{sassum: isolation}
By removing constraints that are not active in the solution set $\cal{A}$ (for which $\lambda_j^* = 0$) from the overall feasible decision set\ \ $\bfs{\mathcal{U}}$ in \nref{def: U}, additional solutions that are connected to $\cal{A}$ are not created. \kraj
\end{stan_assum}{}

We note that, in order to fail this assumption, a quite specific set of conditions must be met. For example, let $F(\bfs{u}) = \col{u_2, -u_1}$, $g_1(\bfs{u}) = a_1 u_1 + b_1 u_2 + c_1$ and $g_2(\bfs{u}) = a_2 u_1 + b_2 u_2 + c_2$. Standard Assumption \ref{sassum: isolation} fails only if  $c_1 = 0$ or $c_2 = 0$ could this assumption fail. Even if Standard Assumption \ref{sassum: isolation} was not satisfied, by Lemma \ref{lemma: unstability} the equilibrium points in $\mathcal{M}\setminus\mathcal{A}$ are unstable, hence there would be no problem in practice.
 \\ \\

Finally, we claim that the dynamics in \nref{eq: full info dynamics} converge to the solutions of the game in \nref{def: game}, if the initial value of the dual variables is different from zero, as formalized next:
\begin{thm}\label{thm: full info dynamics}
Let the Standing Assumptions hold and consider the system dynamics in \nref{eq: full info dynamics}. Then, for any initial condition such that $\omega(0) \notin \mathcal{L}$, there exists a compact set $\mathscr{E} \ni \omega(0)$ which is a superset of $\cal{A}$, such that $\mathcal{A}$ is UGAS for the dynamics restricted to $\mathscr{E}$. \kraj
\end{thm}
\begin{pf}
See Appendix \ref{app: proof theorem 1}.\krajdokaz
\end{pf}
 
\begin{rem}
    Mathematically, it is possible to derive a distributed (center-free) implementation of our semi-decentralized algorithm, similarly to \cite[Equ. 14]{krilavsevic2021learning}, where each agent estimates the dual variables using the information exchanged with the neighbors. While technically possible, this approach is less in line with the almost-decentralized philosophy of extremum seeking, since it would require a dedicated communication network.
\end{rem}

\subsection{Hybrid adaptive gain} \label{sec: hybrid adaptive gain}

\begin{figure}
    \centering
    \includegraphics[width=\linewidth]{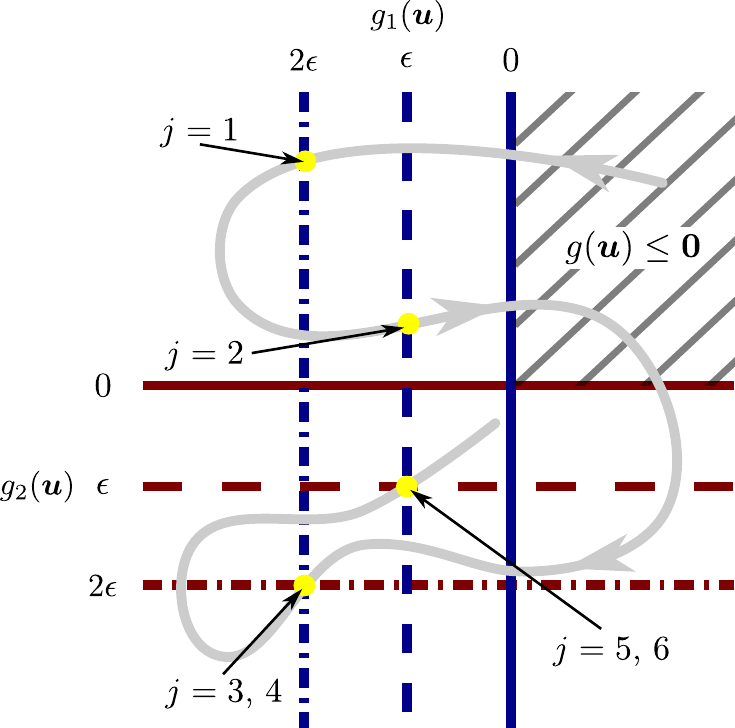}
    \caption{The trajectory is denoted with a gray line, events with yellow dots, first constraint with red and second with blue lines. The trajectory starts in the set where constraints are satisfied ($g(\bfs{u}) \leq 0$). The first event is triggered when the trajectory leaves the set where $g_1(\bfs{u}) \leq 2\epsilon$, causing the state $s_1$ to change to $1$ which then starts the increase of $k_1$ gain. The second event happens when the trajectory returns to the set where $g_1(\bfs{u}) \leq \epsilon$ and the state $s_1$ is reset to 0 which halts the increase in gains. Events 3 and 4 happen when the trajectory leaves the sets $g_1(\bfs{u}) \leq 2\epsilon$ and $g_2(\bfs{u}) \leq 2\epsilon$ simultaneously. In that case, in random order, states $s_1$ and $s_2$ are set to 1. The last jumps happen when the trajectory simultaneously enters the sets $g_1(\bfs{u}) \leq \epsilon$ and $g_2(\bfs{u}) \leq \epsilon$. Again, the states $s_1$ and $s_2$ are reset to 0 in random order. } 
    \label{fig: hybrid switching}
\end{figure}

Due to the properties of the dual dynamics, the coupling constraints can be violated at a certain point in the trajectory. If the cost functions and the constraints are not scaled properly, the pseudogradient can have more influenced than $\nabla g(\bfs{u})$ in the primal dynamics, which in turn would cause the constraints to be active for longer periods of time. When we do not know the cost functions a priori, it is difficult to scale the constraints. To address this potential numerical issue, we propose a gain adaptation scheme based on hybrid dynamical systems, which increases the gains corresponding to violated constraints. The collective flow set and flow map read as:
\begin{subequations}
\begin{align}
    \xi \coloneqq \col{\bfs{u}, \bfs{z}, \lambda, w, k, s} \in C \coloneqq \R^{2m} \times \R_+^{2q}  \times \mathcal{K}^q \times \mathcal{S}^q \label{eq: adaptive gain flow set}\\
  \m{\dot{\bfs{u}} \\  \dot{\bfs{z}} \\ \dot\lambda \\ \dot{{w}} \\ \dot{k} \\ \dot{s}} = F(\xi) \coloneqq \m{-\bfs{u} + \bfs{z} - \Gamma (F(\bfs{u}) + \nabla g(\bfs{u})^\top \lambda ) \\ -\bfs{z} + \bfs{u} \\ \diag{k}\diag{\lambda} \z{g(\bfs{u}) - \lambda + w} \\ -{w} + \lambda \\ \tfrac{1}{2}c(I + S)S^2 \\ \bfs{0}}, \label{eq: adaptive gain flow map}
\end{align} \label{eq: adaptive gain flow}
\end{subequations}
and the collective jump set and jump map
\begin{subequations}
\begin{align}
    &\xi \in D \coloneqq \bigcup_{j = 1}^q D_j,\quad D_j \coloneqq \z{D_j^+ \cup D_j^- \cup D_j^0} \label{eq: adaptive gain jump set}\\
    &\xi^+ \in G(\xi) \coloneqq \Bigg\{ \Big(\bigcup^{j \in \mathcal{C}} G_j(\xi) ,\ \xi \in \bigcap^{j \in \mathcal{C}} D_j \Big)_{\mathcal{C} \in {\mathcal{P}}(\mathcal{Q})}, \label{eq: adaptive gain jump map}
\end{align}\label{eq: adaptive gain jump}
\end{subequations}

where $k$ is a vector of gains for the dual dynamics; $\mathcal{K} \coloneqq [\underline{k}, \overline{k}]$ is the set of possible values for these gains; $s$ is a vector of discrete states which indicate if the gains in $k$ are increasing or not; $\mathcal{S} \coloneqq \{-1, 0, 1\}$ is the set of possible discrete states; $c > 0$ is postive constant which regulates the increase of $k$; $S\coloneqq \diag{s}$, $\epsilon > 0$ is a positive number, $D_j^+ \coloneqq \{\bfs{u} \mid g_j(\bfs{u}) \geq 2\epsilon \} \times \R^m \times \R_+^{2q} \times \mathcal{K}^q \times \mathcal{S}^{j - 1} \times \{-1\} \times \mathcal{S}^{q - j}$ is the set which triggers the increasing $k_j$ dynamics; $D_j^- \coloneqq \{\bfs{u} \mid g_j(\bfs{u}) \leq \epsilon \} \times \R^m \times \R_+^{2q} \times \mathcal{K}^q \times \mathcal{S}^{j - 1} \times \{1\} \times \mathcal{S}^{q - j}$ is the set which triggers the decreasing $k_j$ dynamics; $D_j^0 \coloneqq \R^{2m} \times \R_+^{2q} \times \mathcal{K}^{j - 1} \times \{\overline{k}\} \times \mathcal{K}^{q - j} \times \mathcal{S}^{j - 1} \times \{-1, 1\} \times \mathcal{S}^{q - j}$ is the set which triggers the permanent stop of $k_j$ dynamics; $\mathcal{P}(\mathcal{X})$ is the set of all subsets of $\mathcal{X}$; the jump maps $G_j(\xi)$ are defined as
\begin{align*}
    G_j(\xi) \coloneqq \left\{ \begin{array}{ll}\Delta_{-j} \xi - \Delta_j \xi, & \xi \in D_j^+ \cup D_j^-\\\Delta_{-j} \xi , & \xi \in D_j^0 \\\{\Delta_{-j} \xi - \Delta_j, \Delta_{-j} \xi\}, & \xi \in (D_j^+ \cup   D_j^-) \cap D_j^0\end{array}\right. 
\end{align*}
where $\Delta_{j}$ is a diagonal matrix with all zeros on the diagonal, except for the row corresponding to the $s_j$ state which is equal to one and $\Delta_{-j} \coloneqq I - \Delta_{j}$. \\
In plain words, we have designed an outer-semicontinuous mapping which turns on the increase of the gain $k_j$  when $g_j(\bfs{u}) \geq 2\epsilon$ and turns it off when $g_j(\bfs{u}) \leq \epsilon$ or when the gain reaches the maximum value $\overline{k}$. The set-valued definitions are necessary for outer-semicontinuity, which in turn via hybrid systems theory provides us with some robustness properties. An example trajectory can be seen in Figure \ref{fig: hybrid switching}.
We note that due to the design of the jump sets, no jumps can occur in a sufficiently small neighborhood of a GNE, and no solution can have an infinite number of jumps, as formalized next:
\begin{lem}\label{lemma: number of jumps}
Let the Standing Assumptions hold and let $\xi(t, j)$ be a complete solution to the hybrid system $(C, D, F, G)$ in \nref{eq: adaptive gain flow set}, \nref{eq: adaptive gain flow map}, \nref{eq: adaptive gain jump set} and \nref{eq: adaptive gain jump map}. Then, $\xi(t, j)$ has a finite number of jumps. \kraj
\end{lem}
\begin{pf}
See Appendix \ref{app: proof of jump lemma}. \krajdokaz
\end{pf}
We conclude the section with the convergence result for the proposed hybrid adaptive algorithm.

\begin{thm}\label{thm: adaptive gain dynamics}
 Let the Standing Assumptions hold and consider the hybrid system (C, D, F, G) in \nref{eq: adaptive gain flow set}, \nref{eq: adaptive gain flow map}, \nref{eq: adaptive gain jump set} and \nref{eq: adaptive gain jump map}. Then, for any initial condition such that $\xi(0, 0) \notin \mathcal{L} \times \mathcal{K}^q \times \mathcal{S}^q$ there exists a compact set $\mathscr{K} \supset \mathcal{A} \times \mathcal{K}^q \times \mathcal{S}^q$, such that the set $\mathcal{A} \times \mathcal{K}^q \times \mathcal{S}^q$ is UGAS for the restricted hybrid system $(C \cap \mathscr{K}, D \cap \mathscr{K}, F, G)$. Additionally, the restricted system is structurally robust.\kraj
\end{thm}
\begin{pf}
See Appendix \ref{app: proof of adaptive gain theorem}.\krajdokaz
\end{pf}

\section{Zeroth-order generalized Nash equilibrium seeking} \label{sec: zeroth order GNE}
The main assumptions of Algorithms in $\S$\ref{sec: projectionless gne} and $\S$\ref{sec: hybrid adaptive gain} are that each agent knows their partial-gradient mapping and the actions of other agents. Such knowledge is hard to acquire in practical applications. Our proposed zeroth-order GNE seeking algorithm requires a much weaker assumption; we assume that each agent is only able to measure their cost function. To estimate the pseudogradient via the measurements, we introduce additional oscillator states $\bfs{\mu}$. By injecting oscillations into the inputs of the cost functions, it is possible to  estimate the pseudogradient. For example of a real function of a single variable, it holds that $f(x + a\, \sin(t))\, \sin(t) \approx f(x)\,\sin(t) + a\,\nabla f(x) \sin^2(t)$ for small $a$. If the right-hand expression is averaged in time, only $\tfrac{a}{2} \nabla f(x)$ remains as the desired estimate. The principle is the same for mappings. In order to reduce oscillations, the estimate is then passed through a first-order filter with state $\bfs{\zeta}$ and forwarded into the algorithm in $\S$\ref{sec: hybrid adaptive gain} instead of the real pseudogradient.\\ \\

Our new algorithm is given by
\begin{subequations}
\begin{align}
    &\phi \coloneqq \col{\bfs{u}, \bfs{z}, \lambda, w, k, s, \bfs{\zeta}, \bfs{\mu}} \in C_0 \coloneqq C  \times \R^m \times \mathbb{S}^m \label{eq: adaptive gain zero flow set}\\
    &\m{\dot{\bfs{u}} \\ \dot{\bfs{z}}\\ \dot{\lambda} \\ \dot{{w}} \\ \dot{k} \\ \dot{s} \\ \dot{\bfs{\zeta}} \\ \dot{\bfs{\mu}} } = F_0(\phi) \coloneqq \m{\bfs{\nu} \bfs{\varepsilon}\z{-\bfs{u} + \bfs{z} - \Gamma (\bfs{\zeta} + \nabla g(\bfs{u})^\top \lambda )} \\ \bfs{\nu} \bfs{\varepsilon}\z{-\bfs{z} + \bfs{u}} \\  \nu_0 \varepsilon_0 \diag{k}\diag{\lambda} \z{g(\bfs{u}) - \lambda + w} \\ \nu_0 \varepsilon_0\z{-{w} + \lambda} \\ \tfrac{1}{2}\nu_0 \varepsilon_0 c(I + S)S^2 \\ \bfs{0}\\ \bfs{\nu} \z{- \bfs{\zeta} + \hat F(\bfs{u}, \bfs{\mu})} \\ {2 \pi} \mathcal{R}_{\kappa}\bfs{\mu}}\label{eq: zeroth-order dynamics}
\end{align} \label{eq: zeroth-order algorithm}
\end{subequations}
where $\zeta_i \in \R^{m_{i}}$, $\mu_i \in \mathbb{S}^{m_i}$  are the oscillator states, $\varepsilon_i, \nu_i \geq 0$ for all $i \in \mathcal{I} \cup \{0\}$, $\bfs{\varepsilon} \coloneqq \diag{ (\varepsilon_i I_{m_i})_{i \in \mathcal{I}}}$, $\bfs{\gamma} \coloneqq \diag{ (\gamma_i I_{m_i})_{i \in \mathcal{I}}}$, $\mathcal{R}_{\kappa} \coloneqq \diag{(\mathcal{R}_i)_{i \in \mathcal{I}}}$, $\mathcal{R}_i \coloneqq \diag{(\col{[0, -\kappa_j], [\kappa_j, 0]})_{j \in \mathcal{M}_i}}$, $\kappa_i > 0$ for all $i$ and $\kappa_i \neq \kappa_j$ for $i \neq j$, $\mathcal{M}_j \coloneqq \{\sum_{i = 1}^{j-1}m_i + 1, \dots, \sum_{i = 1}^{j-1}m_i + m_j\}$ is the set of indices corresponding to the state $u_i$ , $\mathbb{D}^n \in \R^{n \times 2n}$ is a matrix that selects every odd row from the vector of size $2n$, $a_i > 0$ are small perturbation amplitude parameters, $A \coloneqq \diag{(a_i I_{m_i})_{j \in \mathcal{I}}}$, $J(\bfs{u}) = \diag{(J_i(u_i, \bfs{u}_{-i})I_{m_i})_{i \in \mathcal{I}}}$,  and $\hat F(\bfs{u}, \bfs{\mu}) = 2A^{-1} J(\bfs{u} + A \mathbb{D}^m \bfs{\mu}) \mathbb{D}^m \bfs{\mu}$. The flow set and map are defined as 
\begin{subequations}
\begin{align}
    &D_0 \coloneqq D \times \R^m \times \mathbb{S}^m \label{eq: adaptive gain zero jump set}\\
    &\phi^+ \in G_0(\phi) \coloneqq \m{G(\xi) \\\xi \\ \mu}. \label{eq: adaptive gain zero jump map}
\end{align}\label{eq: zeroth-order algorithm jumps}
\end{subequations}
Existence of solutions follows directly from \cite[Prop. 6.10]{goebel2009hybrid} as the the continuity of the right-hand side in \nref{eq: zeroth-order algorithm}, \nref{eq: zeroth-order algorithm jumps} and the definitions of flow and jump sets imply \cite[Assum. 6.5]{goebel2009hybrid}. Our main technical result is summarized in the following theorem.
\begin{thm}\label{thm: zeroth order algorithm}

 Let the Standing Assumptions hold and consider the hybrid system $(C_0, D_0, F_0, G_0)$ in \nref{eq: zeroth-order algorithm} and \nref{eq: zeroth-order algorithm jumps}. Then, for any initial condition such that $\phi(0, 0) \notin \mathcal{L} \times \mathcal{K}^q \times \mathcal{S}^q \times \R^m \times \mathbb{S}^m$ there exists a compact set $\mathscr{K} \supset \mathcal{A} \times \mathcal{K}^q \times \mathcal{S}^q \times \R^m \times \mathbb{S}^q$, such that the set $\mathcal{A} \times \mathcal{K}^q \times \mathcal{S}^q \times \R^m \times \mathbb{S}^q$  is SGPAS as $(\bar a, \bar\varepsilon, \bar\nu) = (\max_{i \in \mathcal{I}} a_i, \max_{i \in \mathcal{I}_0}\varepsilon_i, $ $\max_{i \in \mathcal{I}_0}\nu_i) \rightarrow 0$  for the restricted hybrid system $((C \cap \mathscr{K}) \times \R^m \times \mathbb{S}^m), (D \cap \mathscr{K}) \times \R^m \times \mathbb{S}^m, F_0, G_0)$. Additionally, the restricted system is structurally robust.\kraj
\end{thm}

\begin{pf}
See Appendix \ref{app: proof of zeroth order theorem}. \krajdokaz
\end{pf}

\begin{rem}\label{rem: variants}
For the sake of brevity, we made some assumptions with regard to the structure of our proposed algorithms. Namely, we assume that the amplitudes of perturbation signals $a_i$ are constant, that the frequencies of the perturbation signals are different for every state, and that every state of the pseudogradient is estimated. Equivalent results hold for slowly-varying amplitudes $a_i(t) \in [\underline{a}, \overline{a}]$ where the upper and lower bounds are design parameters, for perturbation signals with the same frequency but sufficiently different phases so that ``learning" can occur, and for the pseudogradient with some, but not all, estimated coordinates. \kraj
\end{rem}

\section{Numerical simulations}
\subsection{Two-player monotone game} \label{sec: two player game}
\begin{figure}
    \centering
    \includegraphics{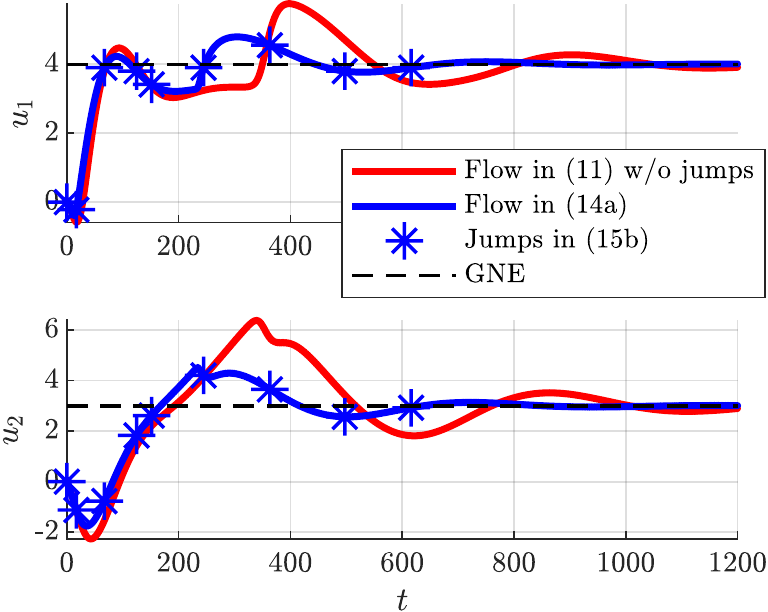}
    \caption{Time evolution of states $u_1$ and $u_2$ for the cases with and without ($k_{\min} = k_{\max}$) adaptive gain in \nref{eq: adaptive gain flow}, \nref{eq: adaptive gain jump map}.}
    \label{fig: exp1_u}
\end{figure}

\begin{figure}
    \centering
    \includegraphics{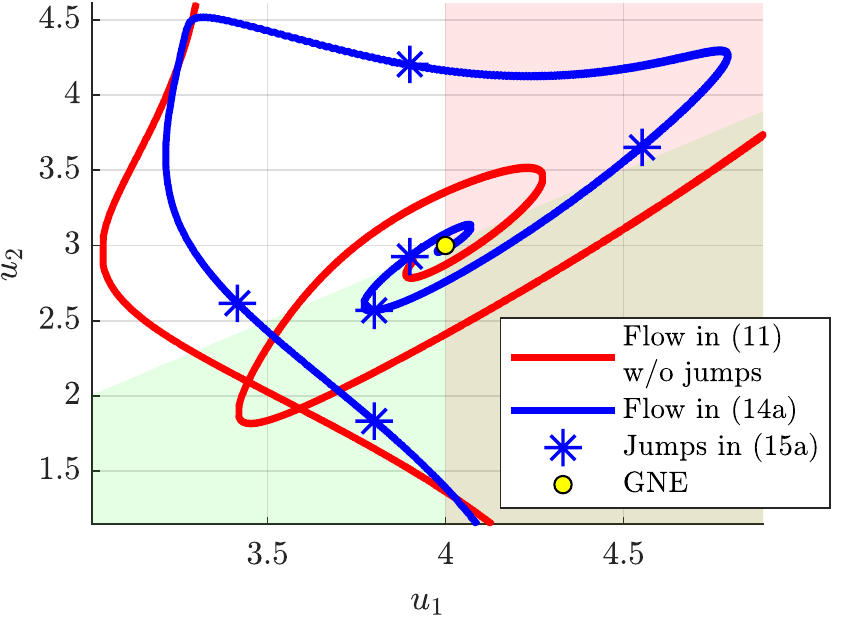}
    \caption{Trajectories with and without adaptive gain in a phase plane. The jumps are activated when entering and leaving the half-spaces corresponding to the constraints (red and green transparent).}
    \label{fig: exp1_phase}
\end{figure}

\begin{figure}
    \centering
    \includegraphics{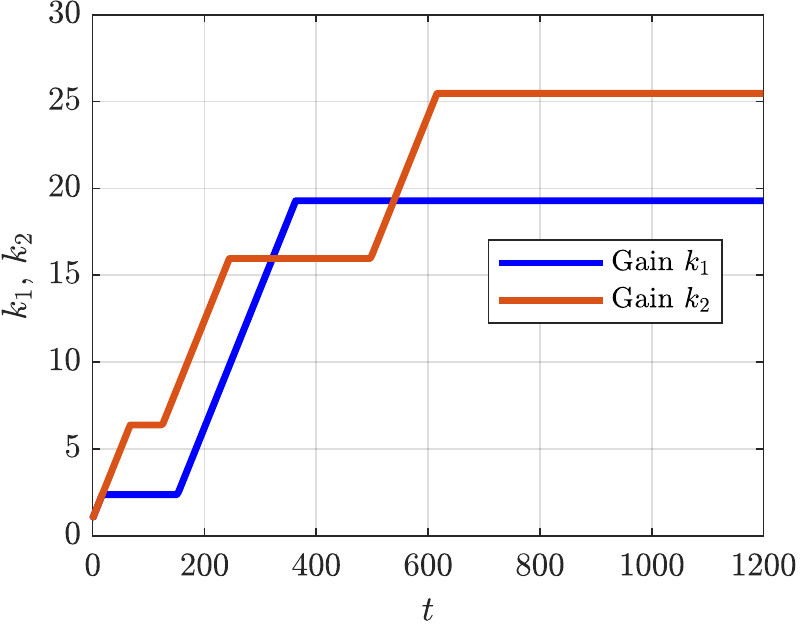}
    \caption{Time evolution of the gains $k_1$ and $k_2$.}
    \label{fig: exp1_gain}
\end{figure}

For our first numerical example, let us consider a two-player monotone game with the following cost functions
\begin{align}
    J_1(\bfs{u}) &= (u_1 - 2)(u_2 + 3) \red
    J_2(\bfs{u}) &= -(u_1 - 2)(u_2 + 3), \label{eq: exp_1 cost}
\end{align}
and constraints
\begin{align}
    u_1 \geq u_2 + 1\text{ and } u_2 \leq 3. \label{eq: exp_1 constraints}
\end{align}
Game in \nref{eq: exp_1 cost} and \nref{eq: exp_1 constraints} has a unique GNE $(u_1^*, u_2^*) = (4, 3)$ and is known to be divergent for algorithms that require strong monotonicity of the pseudogradient. As simulation parameters we choose $a_i = 0.1$, $\varepsilon_i = 0.2$, $\nu_i = 0.2$, $k_j(0, 0) = 1$, $\lambda_j = 0.1$ for all $i \in \mathcal{I}$, $j \in \mathcal{Q}$,  $\epsilon = 0.1$, $k_{\min} = 1$, $k_{\max} = 100$, frequency parameters in range $[11,\, 21]$ and all other initial parameters were set to zero. We compare the algorithm in \nref{eq: adaptive gain flow}, \nref{eq: adaptive gain jump} with algorithm in \nref{eq: full info dynamics} and show the numerical simulations in Figures \ref{fig: exp1_u}, \ref{fig: exp1_phase}. In both cases the algorithm converges to a neighborhood of the GNE, although the convergence is slower in the non-adaptive case. In Figure \ref{fig: exp1_phase}, we denote the area where the constraints are satisfied with green and red rectangles. Jumps correspond to entering the neighborhood of these areas. In Figure \ref{fig: exp1_gain}, we see how the adaptive gain evolves over time.

\subsection{Perturbation signal optimization in oil extraction}
Oil extraction becomes financially unviable when the reservoir pressure drops under a certain threshold. To solve this problem, one can employ gas-lifting. Compressed gas is injected down the well to decrease the density of the fluid and the hydrostatic pressure, causing an increase in production. The oil rate is typically a concave hard-to-model function of the gas injection rate \cite{silva2020dither} with a maximum that slowly changes over time due to changing conditions, making it an excellent candidate for extremum seeking. Extraction sites usually have multiple wells that are operated by the same processing facility. The goal is to maximize the oil extraction rate
\begin{align}
    J_1(\bfs{x}) = \sum_{i = 0}^N f_i(x_i), \label{eq: example_2_cost}
\end{align}
while not exceeding a linear coupling constraint which may relate to total injection rate, power load,  etc.
\begin{align}
    \sum_i^N b_i x_i \leq x_{\max}, \label{eq: example_2_u_constrain}
\end{align}
where $f_i: \R \rightarrow \R$ and $x_i \in \R$ are the oil-rate function and the injection rate, respectively, of the well $i$ and $b_i, x_{\max} \in \R$. We denote the solution to this problem as $x^*$. Furthermore, the processing facility wants to reduce the oscillations in the total optimal extraction rate that result from the extremum seeking perturbation signals:
\begin{align}
    \hat{x}_i(t) = {x}_i(t) + d_i(t) = {x}_i(t) + a_i\, \sin \z{\omega t + \phi_i}.
\end{align}
The oscillations of a single well's optimal extraction rate can be approximated as 
\begin{align*}
    f_i(\hat{x}_i) - f_i(x_i) \approx \nabla f_i(x_i)\, a_i\, \sin \z{\omega t + \phi_i}.
\end{align*}
The secondary goal cannot be accomplished by techniques that diminish the oscillation amplitude over time \cite{abdelgalil2021lie}, \cite{bhattacharjee2021extremum} as the cost functions are slowly-varying and the learning procedure would stop prematurely. Furthermore, we cannot use too high frequencies \cite{suttner2019extremum} as that would also destroy our equipment. Thus, to accomplish our goal, wells are grouped into pairs ($i$, $j$), and each pair selects perturbation signals which are in antiphase:
\begin{align}
    d_i(t) &= a_i\, \sin \z{\omega t + \phi_i} \red
    d_j(t) &= -a_j\, \sin \z{\omega t + \phi_i }. \label{eq: example_2_perturbation_choice}
\end{align}
Without the coupling constraint and with an even number of wells, the perturbation signals in \nref{eq: example_2_perturbation_choice} reduce the oscillations in the neighborhood of the optimum as $\nabla f_1(x_1^*) \approx \nabla f_2(x_2^*) \approx \dots \approx \nabla f_N(x_N^*) \approx 0$. However, if a constraint is present, the perturbation signals might not cancel out properly, because for some pair $(i, j)$ it can hold that $\nabla f_i(x_i^*) \not\approx  \nabla f_j(x_j^*)$. Therefore, it is also necessary to adapt the amplitudes $a_i,\, a_j$ to improve the cancellation effect.  Without loss of generality, we assume that neighboring indices are paired up as in \nref{eq: example_2_perturbation_choice}. The secondary cost function is formulated as follows:
\begin{align}
    \hat{J}_2(a) = &\tfrac{l}{2}\sum_{i=1}^{\frac{N}{2}} \z{ \nabla f_{2i}(x_{2i}^*)\, a_{2i} - \nabla f_{2i - 1}(x_{2i - 1}^*)\, a_{2i - 1}}^2 \red
   & -\sum_{i = 1}^N\log_{p}\z{(a_i - \underline{a})(\overline{a} - a_i)} \nonumber
\end{align}
where $l > 0$, $\underline{a}$ and $\overline{a}$ are the minimum and maximum perturbation amplitude respectively, and it holds $0 < \underline{a} < \overline{a}$. We denote $a^* \coloneqq \argmin\, \hat{J_2}(a)$. Since $x^*$ is not known in advance, direct computation of $a^*$ is not possible. One can modify the previous cost function to use any value of $x$ 
\begin{align}
    {J}_2(x, a) = &\tfrac{l}{2}\sum_{i=1}^{\frac{N}{2}} \z{ \nabla f_{2i}(x_{2i})\, a_{2i} - \nabla f_{2i - 1}(x_{2i - 1})\, a_{2i - 1}}^2, \red
   & -\sum_{i = 1}^N\log_{p}\z{(a_i - \underline{a})(\overline{a} - a_i)}, \label{eq: example_2_cost_perturbation}
\end{align}
and minimize the cost function:
\begin{align}
    & J_p(x, a) = -\alpha J_1(x) + \beta {J}_2(x, a),\quad \alpha,\, \beta > 0,\label{eq: mixed cost}
\end{align}
with constraint \nref{eq: example_2_u_constrain}. However, this approach only approximates the solution $(x^*, a^*)$ for $\alpha \gg \beta$. With our game-theoretic formulation instead, we look for a solution $(x^*, a^*)$ such that $x^*$ is an optimal solution of the oil extraction problem in \nref{eq: example_2_cost} and the overall pair $(x^*, a^*)$ is variational GNE, meaning that the amplitudes are fairly and optimally chosen.\, 

To show that the game is monotone and can be solved by our algorithm, it is sufficient to show that the Jacobian matrix of the pseudogradient is positive semidefinite:

\begin{align}
    \cal{J}_F({x}, {a}) \coloneqq \m{\cal{J}_{11} & \cal{J}_{12} \\ \cal{J}_{21} & \cal{J}_{22}} \succcurlyeq 0.
\end{align}
The submatrix $\cal{J}_{11} \coloneqq \diag{(-\nabla^2 f_i(x_i))_{i \in \cal{I}}}$ is positive semidefinite as all of the cost functions in \nref{eq: example_2_cost} are concave. Furthermore, the submatrix $\cal{J}_{12}$ is a zero matrix as the concave cost functions do not depend on the perturbation amplitudes. Then it holds that $\cal{J}_{22} \coloneqq \diag{\cal{J}_{1,2}, \cal{J}_{3,4}, \dots, \cal{J}_{N - 1, N}}$, where 
\begin{align}
    \cal{J}_{i,j} = l\m{ \z{\nabla f_i(x_i)}^2 & -\nabla f_i(x_i)\nabla f_j(x_j) \\ 
                    -\nabla f_i(x_i)\nabla f_j(x_j) & \z{\nabla f_j(x_j)}^2} \red
                    + \m{\frac{(a_i - \overline{a})^{-2}+ (a_i - \underline{a})^{-2}}{\log(p)} & 0 \\ 0 & \frac{(a_j - \overline{a})^{-2} + (a_j - \underline{a})^{-2}}{\log(p)}}. \label{eq: jacobian ij submatrix}
\end{align}
As both matrices in \nref{eq: jacobian ij submatrix} are positive semidefinite, and $\cal{J}_{22}$ is block diagonal, it follows that the matrix $\cal{J}_{22}$ is positive semidefinite. Finally, due to the block triangular structure of $\cal{J}_F$ and positive semidefinitness of $\cal{J}_{11}$ and $\cal{J}_{22}$, we conclude that $\cal{J}_F$ is positive semidefinite and in turn that the pseudogradient is monotone.\\ \\

\begin{figure}
    \centering
    \includegraphics{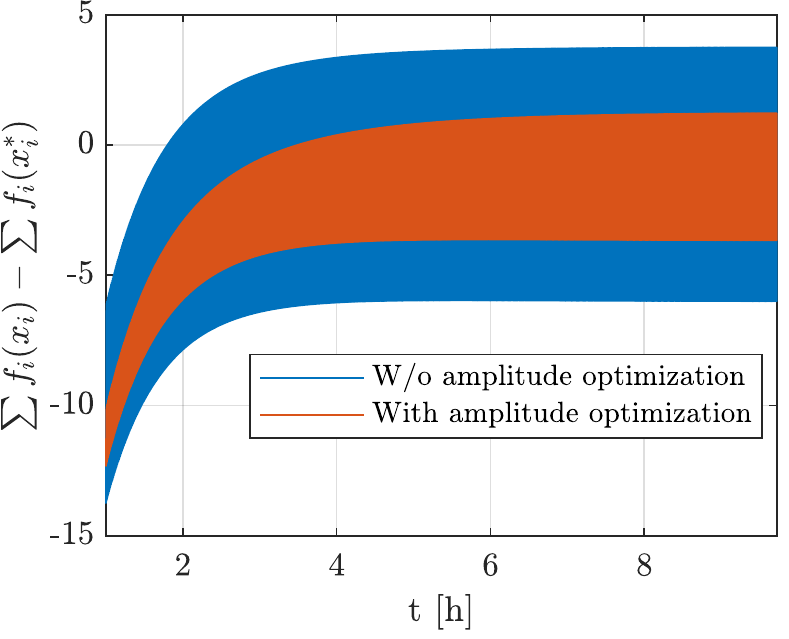}
    \caption{Time evolution of the total oil extraction rate for the case with and without perturbation amplitude optimization.}
    \label{fig:example_2_cost}
\end{figure}

\begin{figure}
    \centering
    \includegraphics{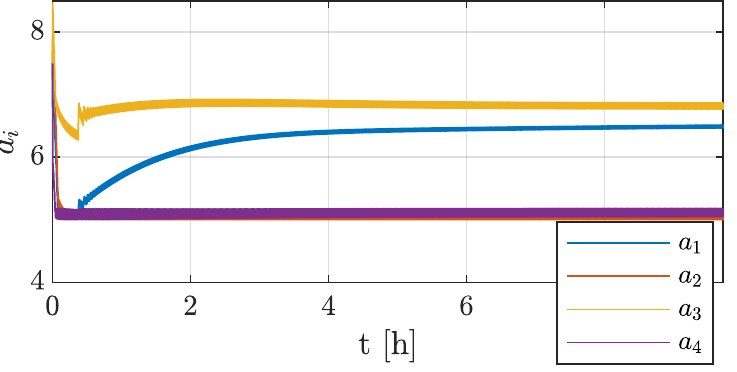}
    \caption{Time evolution of amplitudes $a_i$.}
    \label{fig:example_2_perturbation}
\end{figure}

In our example, the amplitudes of the perturbation signals are part of the decision variable and are therefore time-varying; all perturbation signals have the same frequency but different phases \nref{eq: example_2_perturbation_choice}; and coordinates of the pseudogradient related to cost functions in \nref{eq: example_2_cost_perturbation} need not be estimated, but can be computed directly. Thus, by Remark \ref{rem: variants}, we suitably adjust the algorithm in \nref{eq: zeroth-order algorithm}, \nref{eq: zeroth-order algorithm jumps} and use it for our numerical simulations. Furthermore, we use the well oil extraction rates as in \cite{silva2020dither}

\begin{align*}
f_{1}\left(x_1\right)=&-3.9 \times 10^{-7} x_1^{4}+2.1 \times 10^{-4} x_1^{3} \\
&-0.043 x_1^{2}+3.7 x_1+12, \\
f_{2}\left(x_2\right)=&-1.3 \times 10-7 x_2^{4}+10^{-4} x_2^{3} \\
&-2.8 \times 10^{-2} x_2^{2}+3.1 x_2-17, \\
f_{3}\left(x_3\right)=&-1.2 \times 10^{-7} x_3^{4}+10^{-4} x_3^{3} \\
&-0.028 x_3^{2}+2.5 x_3-16, \\
f_{4}\left(x_4\right)=&-4 \times 10^{-7} x_4^{4}+1.8 \times 10^{-4} x_4^{3} \\
&-0.036 x_4^{2}+3.5 x_4+10, 
\end{align*}

and the following parameters: $l = 10$, $\nu_i = 0.1$, $\varepsilon_i = 0.01$ for all $i$, $\overline{a} = 10$, $\underline{a} = 5$, $p = 100$, $\epsilon = 10$, $\omega_i = 1$, $x_{\max} = 200, b_1 = 1, b_2 = 2, b_3 = 3, b_4 = 4$, $k_{\min} = 10$, $k_{\max} = 10$, $c = 2$, $\Gamma = 10$. For initial conditions: $\bfs{u}(0) = \bfs{z}(0) = \col{10, 10, 10, 10, 7.5, 7.5, 7.5, 7.5}$, $w(0) = 0$, $\lambda(0) = 0.1$, $\zeta(0) = \bfs{0}$, $k(0) = 10$, $s(0) = 0$. Additionally, we run numerical simulations where only the total oil rate is optimized with constant perturbation amplitudes $a_i = 5$, using again the algorithm in \nref{eq: zeroth-order algorithm}. In Figure \ref{fig:example_2_cost}, we see that the amplitude optimization indeed reduces the amplitude of the oscillations in the oil rate by almost 50\% in the steady state, even though larger amplitudes were used in the perturbation signals. In Figure \ref{fig:example_2_perturbation}, we note that in each pair, one of the amplitudes converges to a neighborhood of the minimal value.

\section{Conclusion}
Monotone generalized Nash equilibrium problems with dualized constraints can be solved via a continuous-time golden ratio algorithm augmented by projectionless dual dynamics. Furthermore, the algorithm can be adapted via hybrids systems theory for use in an extremum seeking setting. 
\endlinechar=13
\bibliographystyle{plain}        
\bibliography{biblioteka}  

\endlinechar=-1
\appendix
\section{Proof of Theorem \ref{thm: full info dynamics}} \label{app: proof theorem 1}
We choose the following Lyapunov function candidate 

\begin{align}
&V(\omega, \omega^*) = \tfrac{1}{2}\n{\bfs{u} - \bfs{u}^*}_{\Gamma^{-1}}^2  + \tfrac{1}{2}\n{\bfs{z} - \bfs{z}^*}_{\Gamma^{-1}}^2 \red 
&+ \tfrac{1}{2}\n{w - w^*}^2 + \sum_{j \in \mathcal{Q}}\z{\lambda_j - \lambda_j^* - \lambda_j^* \log\z{\tfrac{\lambda_j}{\lambda_j^*}}}, \label{eq: lyapunov fun cand full info} 
\end{align}
where $\omega^* \in \mathcal{A}$ is any equilibrium point of \nref{eq: full info dynamics} whose $\bfs{u}^*$, $\lambda^*$ states correspond to a GNE and we define $0\, \log 0 \coloneqq 0$. By Standard assumption \ref{sassum: isolation}, equilibrium points in $\cal{M}\setminus\cal{A}$ are disconnected from $\cal{A}$. Furthermore, going back to the Lyapunov function, points in $\cal{M}\setminus\cal{A}$ are not in its domain, and by proving negative semi-definiteness of the Lyapunov derivative, their potential region of attraction is reduced to set $\mathcal{L} \supset \cal{M}$. Thus, we do not consider points in $\cal{L}$ for initial conditions. The Lyapunov derivative is given by
\begin{align}
    \dot{V} &= \vprod{\bfs{u} - \bfs{u}^*}{\Gamma^{-1}\z{-\bfs{u} + \bfs{z} - \Gamma (F(\bfs{u}) + \nabla g(\bfs{u})^\top \lambda )}} \red 
    &\quad + \vprod{\bfs{z} - \bfs{u}^*}{\Gamma^{-1}\z{-\bfs{z} + \bfs{u}}} + \vprod{w - \lambda^*}{-w + \lambda} \red 
    &\quad + \sum_{j \in \mathcal{Q}}\z{\dot{\lambda}_j - \tfrac{\lambda_j^*}{\lambda_j} \dot{\lambda}_j} \red 
    &\leq -\n{\bfs{u} - \bfs{z}}_{\Gamma^{-1}}^2 - \vprod{\bfs{u} - \bfs{u}^*}{ F(\bfs{u}) + \nabla g(\bfs{u})^\top \lambda} \red 
    &\quad \vprod{w - \lambda^*}{-w + \lambda} + { \sum_{j \in \mathcal{Q}}\z{\lambda_j - \lambda_j^*}(g_j(\bfs{u}) -\lambda_j + \omega_j)} \red
    &\leq -\n{\bfs{u} - \bfs{z}}_{\Gamma^{-1}}^2 - \vprod{\bfs{u} - \bfs{u}^*}{ F(\bfs{u}) + \nabla g(\bfs{u})^\top \lambda} \red 
    & \quad + \vprod{w - \lambda^*}{-w + \lambda} + \vprod{\lambda - \lambda^*}{g(\bfs{u}) - \lambda + w} \red
    &\leq -\n{\bfs{u} - \bfs{z}}_{\Gamma^{-1}}^2 - \n{\lambda - w}^2  + \vprod{\lambda - \lambda^*}{g(\bfs{u})}\red
    & \quad +  \vprod{\bfs{u} - \bfs{u}^*}{ F(\bfs{u}) + \nabla g(\bfs{u})^\top \lambda} . \label{eq: lyapunov_ineq_1}
\end{align}

From the properties of v-GNE set, we conclude that
\begin{align}
    \bfs{0}_m &= F\left(\bfs{u}^*\right)+\nabla g(\bfs{u}^*)^{\top}\lambda^* \red
    0 &\leq \vprod{g(\bfs{u}^*)}{\lambda^* - \xi}\text{ for all }\xi \in \R_+^q \label{eq: star properties}
\end{align}
Thus, by using \nref{eq: star properties} within \nref{eq: lyapunov_ineq_1}, we further derive 
\begin{align}
    \dot{V} &\leq -\n{\bfs{u} - \bfs{z}}_{\Gamma^{-1}}^2 - \n{\lambda - w}^2 - \vprod{\bfs{u} - \bfs{u}^*}{ F(\bfs{u}) - F(\bfs{u}^*)} \red
    &\quad - \vprod{\bfs{u} - \bfs{u}^*}{\nabla g(\bfs{u})^\top \lambda - \nabla g(\bfs{u}^*)^\top \lambda^*} \red
    &\quad + \vprod{\lambda - \lambda^*}{g(\bfs{u}) - g(\bfs{u}^*)} \red
    &\leq -\n{\bfs{u} - \bfs{z}}_{\Gamma^{-1}}^2 - \n{\lambda - w}^2 - \underbrace{\vprod{\bfs{u} - \bfs{u}^*}{ F(\bfs{u}) - F(\bfs{u}^*)}}_{\leq 0} \red
    &\quad + \sum_{j \in \mathcal{Q}}\underbrace{\lambda_j}_{\geq 0}\underbrace{\z{ g_j(\bfs{u}) - g_j(\bfs{u}^*) + \vprod{\bfs{u}^* - \bfs{u}}{\nabla g_j(\bfs{u})}}}_{\leq 0} \red
    &\quad - \sum_{j \in \mathcal{Q}}\underbrace{\lambda_j^*}_{\geq 0}\underbrace{\z{g_j(\bfs{u}) - g_j(\bfs{u}^*) -\vprod{\bfs{u} - \bfs{u}^*}{\nabla g_j(\bfs{u}^*)}}}_{\geq 0} \red
    &\leq -\n{\bfs{u} - \bfs{z}}_{\Gamma^{-1}}^2 - \n{\lambda - w}^2, \label{eq: lyapunov_ineq_2}
\end{align}
where the last inequality follows from the monotonicity of the pseudogradient and the convexity of the coupled constraints. Now, we prove via La Salle's theorem that the trajectories of \nref{eq: full info dynamics} converge to the set $\mathcal{A}$. Let us define the following sets:
\begin{align}
    \Omega_c &\coloneqq \{\omega \in \R^{2m} \times \R_+^{2q} \mid V(\omega) \leq c\} \red
    \Omega_0 &\coloneqq \{\omega \in \Omega_c \mid \bfs{u} = \bfs{z} \text{ and } \lambda = w\} \red
    \mathcal{Z} &\coloneqq \{\omega \in \Omega_c \mid \dot{V}(\omega) = 0\} \red
    \mathcal{O} &\coloneqq \{\omega \in \Omega_c \mid \omega(0) \in \mathcal{Z} \Rightarrow  \omega(t) \in \mathcal{Z}\  \forall t \in \R\}, 
\end{align}
where $\Omega_c$ is a non-empty compact sublevel set of the Lyapunov function candidate, $\mathcal{Z}$ is the set of zeros of its derivative, $\Omega_0$ is the superset of $\mathcal{Z}$ which follows from \nref{eq: lyapunov_ineq_2} and $\mathcal{O}$ is the maximum invariant set as explained in \cite[Chp. 4.2]{khalil2002nonlinear}. Then, for some $c>0$ large enough, it holds that
\begin{align}
    \Omega_c \supseteq \Omega_0 \supseteq \mathcal{Z} \supseteq \mathcal{O} \supseteq \mathcal{A}.
\end{align}
Firstly, for any compact set $\Omega_c$, since the right-hand side of \nref{eq: full info dynamics} is (locally) Lipschitz continuous and therefore by \cite[Thm. 3.3]{khalil2002nonlinear} we conclude that solutions to \nref{eq: full info dynamics} exist and are unique. Next, we show that the only $\omega$-limit trajectories in $\mathcal{O}$ are the equilibrium points of the dynamics in \nref{eq: full info dynamics}, i.e. $\mathcal{O} \equiv \mathcal{A}$. It is sufficient to prove that there cannot exist any positively invariant trajectories in $\Omega_0$, apart from stationary points in $\mathcal{A}$. For trajectories in $\Omega_0$, it holds that
\begin{align}
    \bfs{0} &= \bfs{u} - \bfs{z} \label{eq: razlika u z}\\
    \bfs{0} &= \dot{\bfs{u}} - \dot{\bfs{z}} \label{eq: razlika izvoda u z}\\
    \bfs{0} &= \lambda - w \label{eq: razlika lambda w}\\
    \bfs{0} &= \dot{\lambda} - \dot{w}, \label{eq: razlika izvod lambda w}
\end{align}
and therefore
\begin{align}
    \bfs{0} &= F\left(\bfs{u}\right)+\nabla g(\bfs{u})^{\top} \lambda \label{eq: omega_0 primal}\\
    \bfs{0} &= \diag{\lambda} g(\bfs{u}), \label{eq: omega_0 dual}
\end{align}

where \nref{eq: omega_0 primal} follows from \nref{eq: full info dynamics} and \nref{eq: razlika izvoda u z}, and \nref{eq: omega_0 dual} follows from \nref{eq: full info dynamics}, \nref{eq: razlika lambda w} and \nref{eq: razlika izvod lambda w}. Equations  \nref{eq: razlika izvoda u z}, \nref{eq: razlika lambda w}, \nref{eq: omega_0 primal} and \nref{eq: omega_0 dual} form the definition of set $\mathcal{M}$ in \nref{eq: equilibrium points} and the fact that $\cal{M}\setminus\cal{A}$ is not in the domain, we conclude $\Omega_0 \equiv \mathcal{A}$. Since the set $\mathcal{O}$ is a subset of the set $\Omega_0$, we conclude that $\mathcal{O} \equiv \mathcal{A}$. Therefore, by La Salle's theorem \cite[Thm. 4]{khalil2002nonlinear}, set $\mathcal{A}$ is attractive for the dynamics in \nref{eq: full info dynamics}. \\
Next, we prove stability of $\mathcal{A}$. We restrict the domain of the dynamics by choosing an arbitrary $\omega^*$ and set a $\Lambda$ that contains arbitrarily many initial conditions of interest not contained in the set $\mathcal{L}$,  and it holds $\mathcal{A} \subset \Lambda$. Then, we compute $\overline{c} = \max_{\omega \in \Lambda} V(\omega, \omega^*)$ and define the new restricted domain to the forward invariant set $\mathscr{E}$, where $\mathscr{E} \coloneqq \{\omega \in \R^{2m} \times \R_+^{2q} \mid V(\omega, \omega^*) \leq \overline{c}\}$.\\
Consequently, we define the following set-valued mapping of compact sets $\Omega(\omega^*, c) \coloneqq \{\omega \in \mathscr{E}\mid V(\omega, \omega^*) \leq c\}$. Now, we show global stability with respect to the set $\mathcal{A}$. Let us choose an arbitrary $\varepsilon > 0$. For a particular $c$ and $\omega^*$, since $V$ does not increase, it follows that all trajectories that start in $\Omega(\omega^*, c)$ are contained in the set. Let us choose $c(\omega^*)$ such that $\Omega(\omega^*, c(\omega^*)) \subseteq (\mathcal{A} + \varepsilon \mathbb{B}) \cap \mathscr{E}$. By continuity of $V$, for every set $\Omega(\omega^*, c(\omega^*))$, it is possible to find $\delta(\omega^*) > 0$ such that $(\omega^* + \delta(\omega^*)\mathbb{B})\cap \mathscr{E} \subseteq \Omega(\omega^*, c(\omega^*))$. If we take $\delta = \min_{\omega^* \in \mathcal{A}} \delta(\omega^*)$, it holds that $\cup_{\omega^* \in \mathcal{A}} (\omega^* + \delta\mathbb{B})\cap \mathscr{E} = (\mathcal{A} + \delta \mathbb{B})\cap \mathscr{E}$. Thus, $(\mathcal{A} + \delta \mathbb{B})\cap \mathscr{E} \subseteq \cup_{\omega^* \in \mathcal{A}} \Omega(\omega^*, c(\omega^*))$ which implies that all solutions with $\omega(0) \in (\mathcal{A} + \delta \mathbb{B})$, remain in $(\mathcal{A} + \varepsilon \mathbb{B})$ for all $t\geq 0$. Therefore, set $\mathcal{A}$ is globally stable and attractive on $\mathscr{E}$, hence it is UGAS.

\section{Proof of Lemma 1} \label{app: proof of lemma 1}
We study the stability of singular equilibrium points in the set $\cal{M}\setminus\mathcal{A}$. The main difference between the set $\mathcal{M}\setminus\cal{A}$ and the set of solutions $\mathcal{A}$, is that the set $\mathcal{M}\setminus\cal{A}$ can contain points where $\bar{\lambda}_j = 0$ and $g_i(\bar{\bfs{u}}) > 0$ for some index $j$. Let $\hat{\omega} \in \mathcal{M}\setminus\cal{A}$. Without loss of generality, we assume that for $j = q$ it holds that $\hat{\lambda}_q = 0$ and $g_q(\hat{\bfs{u}}) > 0$. In order to check the stability of the point $\hat{\omega}$, we study the dynamics in \nref{eq: full info dynamics} linearized around $\hat{\omega}$:


\begin{align}
    \m{\dot{\Tilde{\bfs{z}}} \\ \dot{\Tilde{\bfs{u}}} \\ \dot{\Tilde{w}} \\ \dot{\Tilde{\lambda}}} = \left[\begin{array}{c c c |c} 
        -I_m & I_m & \bfs{0} & \bfs{0} \\ I_m & -I_m - M & \bfs{0} & -\nabla g(\hat{\bfs{u}})^\top \\  \bfs{0} & \bfs{0} & -I_q & I_q \\ 0 & 0 & 0 & g_1(\hat{\bfs{u}}) \\ \vdots & \vdots & \vdots & \vdots \\  \hline 0 & 0 & 0 & g_q(\hat{\bfs{u}})
    \end{array}\right]  \m{{\Tilde{\bfs{z}}} \\ {\Tilde{\bfs{u}}} \\ {\Tilde{w}} \\ {\Tilde{\lambda}}},
\end{align}
where $\Tilde{\bfs{z}} \coloneqq {\bfs{z}} - \hat{\bfs{z}}$, $\Tilde{\bfs{u}} \coloneqq {\bfs{u}} - \hat{\bfs{u}}$, $\Tilde{w} \coloneqq {w} - \hat{w}$, $\Tilde{\lambda} \coloneqq {\lambda} - \hat{\lambda}$ and $M(\hat{\bfs{u}}, \hat{\lambda}) \coloneqq \left. \frac{\partial}{\partial \bfs{u}}\z{\Gamma (F(\bfs{u}) + \nabla g(\bfs{u})^\top \lambda )} \right|_{\bfs{u} = \hat{\bfs{u}}, \lambda = \hat{\lambda}}$. The system matrix will have at least one positive eigenvalue due to the upper triangular structure and the element $g_q(\hat{\bfs{u}}) > 0$ in the last row. It follows that the equilibrium point $\hat{\omega}$ is unstable for dynamics in \nref{eq: full info dynamics}. As $\hat{\omega}$ was chosen arbitrarily, we conclude that any equilibrium point in $\mathcal{M} \setminus \mathcal{A}$ is unstable.

\section{Proof of Lemma \ref{lemma: number of jumps}} \label{app: proof of jump lemma}
Let us assume otherwise, that we have an infinite amount of jumps. By the structure of the jump set and map, we must jump between $s_i = -1$ and $s_i = 1$ an infinite amount of times for at least one of the states $i$. Without the loss of generality, we assume that this is true for $i = j$. As we can spend only a finite amount of time in the state $s_j = 1$ ($\tau = \frac{\overline{k} - \underline{k}}{c_j}$), time between jumps from $s_j = 1$ to $s_j = -1$, $t_k$, has to decrease to zero, otherwise $\sum^\infty t_k = \infty > \tau$. Minimum time between jumps $t_{\min}$ is equal to $\frac{d_{\min}}{\max{\n{\dot{\bfs{u}}}}}$, where $d_{\min}$ is the minimal distance between the jump sets corresponding to $s_j = -1$ and $s_j = 1$, which exists by Lemma \ref{lem: minimum distance} and is positive, and $\max{\n{\dot{\bfs{u}}}}$ is finite based on the continuity of the flow map and the forward invariance of any compact set $\Omega_c$. As both are finite positive numbers, we conclude that $t_{\min} > 0$, which leads us to a contradiction. Therefore, we can only have a finite number of jumps.

\begin{lem}\label{lem: minimum distance} Let $G_j(\epsilon) \coloneqq \{\bfs{y} \mid g_j(\bfs{y}) = \epsilon\}$. For any convex constraint $g_j(\bfs{u})$, the Euclidean distance $d_{\min} \coloneqq \min_{(\bfs{u}, \bfs{v}) \in G_j(\epsilon) \times G_j(2\epsilon)}\n{\bfs{u} - \bfs{v}}$ for arbitrarily small $\epsilon > 0$ exists and it is larger than zero. \kraj
\end{lem}
\begin{pf}
Let us choose $\epsilon$ such that $G_j(2\epsilon) \neq \emptyset$. By convexity property of the constraint function, for $\bfs{u} \in G_j(\epsilon)$ and $\bfs{v} \in G_j(2\epsilon)$, we have:
\begin{align*}
    g_j(\bfs{u}) &\geq g_j(\bfs{v}) + \nabla g_j(\bfs{v}) (\bfs{u} - \bfs{v}) \\
    \epsilon &\leq \nabla g_j(\bfs{v}) (\bfs{v} - \bfs{u}) \leq \n{\nabla g_j(\bfs{v})}\n{\bfs{v} - \bfs{u}}.\\
    \frac{\epsilon}{\n{\nabla g_j(\bfs{v})}} &\leq \n{\bfs{v} - \bfs{u}}
\end{align*}
As the set $G_j(2\epsilon)$ is compact, and $\nabla g_j(\bfs{v})$ is continuous in its coordinates, by the extreme value theorem, $\n{\nabla g_j(\bfs{v})}$ reaches a maximum $\delta$ on that set. Therefore, the minimum distance is bounded bellow as $d_{\min} \geq \frac{\epsilon}{\delta}$.
\krajdokaz
\end{pf}
\section{Proof of Theorem \ref{thm: adaptive gain dynamics}}\label{app: proof of adaptive gain theorem}
Proof of convergence is similar to that of Theorem \ref{thm: full info dynamics}. First, we note that the additional states are invariant to the set $\mathcal{K}^q\times\mathcal{S}^q$ regardless of the rest of the dynamics. Next, we choose the Lyapunov function candidate
\begin{align}
&V(\omega, \omega^*, k) = \tfrac{1}{2}\n{\bfs{u} - \bfs{u}^*}_{\Gamma^{-1}}^2  + \tfrac{1}{2}\n{\bfs{z} - \bfs{u}^*}_{\Gamma^{-1}}^2 \red 
&+ \tfrac{1}{2}\n{w - \lambda^*}^2  + \sum_{j \in \mathcal{Q}}\tfrac{1}{k_j}\z{\lambda_j - \lambda_j^* - \lambda_j^* \log\z{\tfrac{\lambda_j}{\lambda_j^*}}},
\end{align}
which depends on the chosen equilibrium point $\omega^*$. In a similar manner as in the proof of Theorem \ref{thm: full info dynamics}, it follows that
\begin{align}
    u_c(\xi) &= \vprod{\nabla V(\xi)}{F(\xi)} \leq -\n{\bfs{u} - \bfs{z}}_{\Gamma^{-1}}^2 - \n{\lambda - w}^2, \label{eq: invariant flow}\\
    u_d(\xi) &= V(\omega_+, \omega^*, k) - V(\omega, \omega^*, k) = 0. \label{eq: invariant jump}
\end{align}

We restrict the flow and jump sets by choosing an arbitrary $\omega^*$ and set $\Lambda$ that contains arbitrarily many initial conditions of interest not contained in the set $\mathcal{L}$, and it holds $\mathcal{A} \subset \Lambda$. Then, we compute $\overline{c} = \max_{\omega \in \Lambda} V(\omega, \omega^*, k_{\max})$ and define the new restricted flow set as $\mathscr{K} \coloneqq \mathscr{E} \times \mathcal{K}^q \times \mathcal{S}^q$, where $\mathscr{E} \coloneqq \{\omega \in \R^{2m} \times \R_+^{2q} \mid V(\omega, \omega^*, k_{\min}) \leq \overline{c}\}$.\\

Consequently, we define the following set-valued mapping of compact sets $\Omega(\omega^*, k, c) \coloneqq \{\omega \in \mathscr{E}\mid V(\omega, \omega^*, k) \leq c\}$. It holds that $\bfs{0} < k' \leq k''$ implies that $\Omega(\omega^*, k', c) \subseteq \Omega(\omega^*, k'', c)$. As $k$ is dynamic, the ``invariant set'', in which the trajectories of $\omega$ dynamics are contained, expands in the $\lambda$ dimensions. \\

Now, we show global stability with respect to the set $\mathcal{A}\times\mathcal{K}^q\times \mathcal{S}^q$. Let us choose an arbitrary $\varepsilon > 0$. For a particular $c$ and $\omega^*$, the trajectories are constrained to the largest $\Omega$ set for $k = k_{\max}$, and to the smallest for $k = k_{\min}$. Therefore, by the fact that $V$ does not increase during flows or jumps, and that the gains $k$ are constrained to the set $\mathcal{K}^q$, it follows that all trajectories that start in $\Omega(\omega^*, k_{\min}, c)$ are contained in the set $\Omega(\omega^*, k_{\max}, c)$. Let us choose $c(\omega^*)$ such that $\Omega(\omega^*, k_{\max}, c(\omega^*)) \subseteq (\mathcal{A} + \varepsilon \mathbb{B}) \cap \mathscr{E}$. By continuity of $V$, for every set $\Omega(\omega^*, k_{\min}, c(\omega^*))$, it is possible to find $\delta(\omega^*) > 0$ such that $(\omega^* + \delta(\omega^*)\mathbb{B})\cap \mathscr{E} \subseteq \Omega(\omega^*, k_{\min}, c(\omega^*))$. If we take $\delta = \min_{\omega^* \in \mathcal{A}} \delta(\omega^*)$, it holds that $\cup_{\omega^* \in \mathcal{A}} (\omega^* + \delta\mathbb{B})\cap \mathscr{E} = (\mathcal{A} + \delta \mathbb{B})\cap \mathscr{E}$. Thus, $(\mathcal{A} + \delta \mathbb{B})\cap \mathscr{E} \subseteq \cup_{\omega^* \in \mathcal{A}} \Omega(\omega^*, k_{\min}, c(\omega^*))$ which implies that all maximal solutions with $\xi(0, 0) \in (\mathcal{A} + \delta \mathbb{B}) \times \mathcal{K}^q \times \mathcal{S}^q$, remain in $(\mathcal{A} + \varepsilon \mathbb{B}) \times \mathcal{K}^q \times \mathcal{S}^q$ for all $(t, j) \in \dom \xi$. Next, we prove global pre-attractivity for the constrained flow and jump sets. Let $\xi$ be a complete solution in $\mathscr{K}$. For a fixed $\omega^*$, we define $\hat{V}(\xi)\coloneqq V(\omega, \omega^*, k)$. Via \cite[Cor. 8.7]{goebel2012hybrid} and Lemma \ref{lemma: number of jumps}, we conclude that for some $r \geq 0$, $\xi$ approaches the largest weakly invariant subset in $\hat{V}^{-1}(r) \cap \mathscr{K} \cap\  \overline{u_{c}^{-1}(0)}$,
where the notation $f^{-1}(r)$ stands for the $r$-level set of $f$ on $\operatorname{dom}f$, the domain of definition of $f$, i.e., $f^{-1}(r):=\{z \in \operatorname{dom} f \mid f(z)=r\}$. By same reasoning as in Theorem \ref{thm: full info dynamics}, we conclude that $\overline{u_{c}^{-1}(0)} = \mathcal{A}  \times \mathcal{K}^q \times \mathcal{S}^q$. Thus, the largest weakly invariant subset for $\xi$ reads as $ \hat{V}^{-1}(r) \cap \z{\mathcal{A} \times \mathcal{K}^q \times \mathcal{S}^q}$. Every trajectory $\xi$ converges to a different subset. The union of invariant subsets for every trajectory is $\mathcal{A} \times \mathcal{K}^q \times \mathcal{S}^q$, as we can choose an initial condition for which it holds $\omega(0, 0) = \omega^* = const.$ for all $(t, j) \in \dom \xi$, for any $\omega^* \in \mathcal{A}$. Therefore, $\mathcal{A} \times \mathcal{K}^q \times \mathcal{S}^q$ is globally attractive, as all solutions are complete, which implies that the set $\mathcal{A} \times \mathcal{K}^q \times \mathcal{S}^q$ is UGAS (\cite[Thm. 7.12]{goebel2012hybrid}) on the restricted flow and jump sets. Furthermore, by \cite[Prop. A.1.]{poveda2021robust}, the HDS $(C \cap \mathscr{K}, D \cap \mathscr{K}, F, G)$ is structurally robust.

\section{Proof of Theorem \ref{thm: zeroth order algorithm}}\label{app: proof of zeroth order theorem}
We rewrite the system in \nref{eq: zeroth-order dynamics} as 

\begin{align}
     \m{\dot{\bfs{u}} \\ \dot{\bfs{z}} \\ \dot{\lambda} \\ \dot{{w}} \\ \dot{k} \\ \dot{s} \\ \dot{\bfs{\zeta}}} &=
     \m{ \bar\nu \bar\varepsilon \tilde{\bfs{\nu}} \tilde{\bfs{\varepsilon}}\z{-\bfs{u} + \bfs{z} - \Gamma (\bfs{\zeta} + \nabla g(\bfs{u})^\top \lambda )} \\ 
     \bar\nu \bar\varepsilon \tilde{\bfs{\nu}} \tilde{\bfs{\varepsilon}}\z{-\bfs{z} + \bfs{u}} \\ 
     \bar\nu \bar\varepsilon \tilde{\nu}_0 \tilde{\varepsilon}_0 \diag{k}\diag{\lambda} \z{g(\bfs{u}) - \lambda + w} \\
     \bar\nu \bar\varepsilon \tilde{\nu}_0 \tilde{\varepsilon}_0\z{-{w} + \lambda} \\ 
     \tfrac{1}{2}\bar\nu \bar\varepsilon \tilde{\nu}_0 \tilde{\varepsilon}_0 c(I + S)S^2 \\ 
     \bfs{0}\\
      \bar{\nu}\tilde{\bfs{\nu}} \z{- \bfs{\zeta} + \hat F(\bfs{u}, \bfs{\mu})}},\label{eq: zeroth order dynamics rewriten} \\
     \dot{\bfs{\mu}} &=  {2 \pi} \mathcal{R}_{\kappa}\bfs{\mu}, \label{eq: oscilator}
\end{align}

where $\bar\nu \coloneqq \max_{i \in \mathcal{I}_0} \nu_i$, $\bar\varepsilon \coloneqq \max_{i \in \mathcal{I}_0} \varepsilon_i$, $\tilde{\bfs{\nu}} \coloneqq \bfs{\nu}/{\bar \nu}$, $\tilde{\bfs{\varepsilon}} \coloneqq \bfs{\varepsilon}/{\bar \varepsilon}$, $\tilde{{\nu}_0} \coloneqq {\nu_0}/{\bar \nu}$ and $\tilde{{\varepsilon}_0} \coloneqq {\varepsilon_0}/{\bar \varepsilon}$.  The system in \nref{eq: zeroth order dynamics rewriten}, \nref{eq: oscilator} is in singular perturbation from where $\bar \nu$ is the time scale separation constant. The goal is to average the dynamics of $\xi, \bfs{\zeta}$ along the solutions of $\bfs{\mu}$. For sufficiently small $\bar a \coloneqq \max_{i \in \mathcal{I}} a_i$, we can use the Taylor expansion to write down the cost functions as 
\begin{align}
    &J_i(\bfs{u} + A \mathbb{D} \bfs{\mu}) = J_i(u_i, \bfs{u}_{-i}) + a_i (\mathbb{D}^{m_i} \mu_i)^\top \nabla_{u_i}J_i(u_i, \bfs{u}_{-i}) \red 
    &+ A_{-i} (\mathbb{D}^{m_{-i}} \bfs{\mu}_{-i})^\top \nabla_{u_{-i}}J(u_i, \bfs{u}_{-i}) + O(\bar a^2), \label{eq: Taylor approx}
\end{align}
where $A_{-i} \coloneqq \diag{(a_i I_{m_i})_{j \in \mathcal{I} \setminus \{i\}}}$. By the fact that the right-hand side of \nref{eq: zeroth order dynamics rewriten}, \nref{eq: oscilator} is continuous, by using \cite[Lemma 1]{poveda2020fixed} and by substituting \nref{eq: Taylor approx} into \nref{eq: zeroth order dynamics rewriten}, we derive the well-defined average of the complete dynamics:
\begin{align}
     \m{\dot{\bfs{u}} \\ \dot{\bfs{z}} \\ \dot{\lambda} \\ \dot{{w}} \\ \dot{k} \\ \dot{s} \\ \dot{\bfs{\zeta}}} &=
     \m{\bar\varepsilon \tilde{\bfs{\nu}} \tilde{\bfs{\varepsilon}}\z{-\bfs{u} + \bfs{z} - \Gamma (\bfs{\zeta} + \nabla g(\bfs{u})^\top \lambda )} \\
     \bar\varepsilon \tilde{\bfs{\nu}} \tilde{\bfs{\varepsilon}}\z{-\bfs{z} + \bfs{u}} \\ 
     \bar\varepsilon \tilde{\nu}_0 \tilde{\varepsilon}_0 \diag{\lambda} \z{g(\bfs{u}) - \lambda + w} \\
      \bar\varepsilon \tilde{\nu}_0 \tilde{\varepsilon}_0\z{-{w} + \lambda} \\ 
      \tfrac{1}{2}\bar\varepsilon \tilde{\nu}_0 \tilde{\varepsilon}_0 c(I + S)S^2 \\ 
     \bfs{0}\\
      \tilde{\bfs{\nu}} \z{- \bfs{\zeta} + F(\bfs{u}) + \mathcal{O}(\bar a)}} .\label{eq: average dynamics with O(a)}
\end{align}
The system in \nref{eq: average dynamics with O(a)} is an $\mathcal{O}(\bar a)$ perturbed version of:
\begin{align}
     \m{\dot{\bfs{z}} \\ \dot{\bfs{u}} \\ \dot{{w}} \\ \dot{\lambda}\\ \dot{k} \\ \dot{s}  \\ \dot{\bfs{\zeta}}} &=
     \m{\bar\varepsilon \tilde{\bfs{\nu}} \tilde{\bfs{\varepsilon}}\z{-\bfs{z} + \bfs{u}} \\ 
      \bar\varepsilon \tilde{\bfs{\nu}} \tilde{\bfs{\varepsilon}}\z{-\bfs{u} + \bfs{z} - \Gamma (\bfs{\zeta} + \nabla g(\bfs{u})^\top \lambda )} \\
      \bar\varepsilon \tilde{\nu}_0 \tilde{\varepsilon}_0\z{-{w} + \lambda} \\ 
     \bar\varepsilon \tilde{\nu}_0 \tilde{\varepsilon}_0 \diag{\lambda} \z{g(\bfs{u}) - \lambda + w} \\
      \tfrac{1}{2}\bar\varepsilon \tilde{\nu}_0 \tilde{\varepsilon}_0 c(I + S)S^2 \\ 
     \bfs{0}\\
     \tilde{\bfs{\nu}} \z{- \bfs{\zeta} + F(\bfs{u})}} .\label{eq: real nominal average dynamics}
\end{align}
For sufficiently small $\bar \varepsilon$, the system in \nref{eq: real nominal average dynamics} is in singular perturbation form with dynamics $\bfs{\zeta}$ acting as fast dynamics. The boundary layer dynamics are given by

\begin{align}
    \dot{\bfs{\zeta}}_{\text{bl}} = \tilde{\bfs{\nu}} \left( - \bfs{\zeta}_{\text{bl}} +  F(\bfs{u}_{\text{bl}})\right) \label{boundary layer dynamics}
\end{align}

For each fixed $\bfs{u}_{\text{bl}}$, $\{F(\bfs{u}_{\text{bl}})\}$ is an uniformly globally exponentially stable equilibrium point of the boundary layer dynamics. By \cite[Exm. 1]{wang2012analysis}, it holds that the system in \nref{eq: real nominal average dynamics} has a well-defined average system given by

\begin{align}
     \m{\dot{\bfs{z}} \\ \dot{\bfs{u}} \\ \dot{{w}} \\ \dot{\lambda} \\ \dot{k} \\ \dot{s}} &=
     \m{\tilde{\bfs{\nu}} \tilde{\bfs{\varepsilon}}\z{-\bfs{z} + \bfs{u}} \\ 
     \tilde{\bfs{\nu}} \tilde{\bfs{\varepsilon}}\z{-\bfs{u} + \bfs{z} - \Gamma (F(\bfs{u}) + \nabla g(\bfs{u})^\top \lambda )} \\
     \tilde{\nu}_0 \tilde{\varepsilon}_0\z{-{w} + \lambda} \\ 
     \tilde{\nu}_0 \tilde{\varepsilon}_0 \diag{\lambda} \z{g(\bfs{u}) - \lambda + w} \\
     \tfrac{1}{2}\tilde{\nu}_0 \tilde{\varepsilon}_0 c(I + S)S^2 \\ 
     \bfs{0}\\}.\label{eq: reduced system dynamics dynamics}
\end{align}
To prove that the system in \nref{eq: reduced system dynamics dynamics} renders the set $\mathcal{A} \times \mathcal{K}^q \times \mathcal{S}^q$ UGAS for restricted dynamics, we consider the following Lyapunov function candidate:
\begin{align}
    &V(\xi, \omega^*) = \tfrac{1}{2}\n{\bfs{u} - \bfs{u}^*}_{{\z{\tilde{\bfs{\nu}} \tilde{\bfs{\varepsilon}}\Gamma}}^{-1}}^2  + \tfrac{1}{2}\n{\bfs{z} - \bfs{u}^*}_{{\z{\tilde{\bfs{\nu}} \tilde{\bfs{\varepsilon}}\Gamma}}^{-1}}^2 \red 
    &+\tfrac{1}{2\tilde{\nu}_0 \tilde{\varepsilon}_0}\n{w - \lambda^*}^2 + \sum_{j \in \mathcal{Q}}\tfrac{1}{\tilde{\nu}_0 \tilde{\varepsilon}_0 k_j}\z{\lambda_j - \lambda_j^* - \lambda_j^* \log\z{\tfrac{\lambda_j}{\lambda_j^*}}}.
    \label{eq: lyapunov fun cand}
\end{align}
The convergence proof is equivalent to the proof of Theorem \ref{thm: adaptive gain dynamics} and is omitted. We restrict the flow and jump sets as $C \cap \mathscr{K}$ and $D \cap \mathscr{K}$ respectively.\\ \\

Next, by \cite[Thm. 2, Exm. 1]{wang2012analysis}, the dynamics in \nref{eq: real nominal average dynamics} render the set $\mathcal{A} \times \mathcal{K}^q \times \mathcal{S}^q \times \R^m$ SGPAS as $(\bar \varepsilon \rightarrow 0)$. As the right-hand side of the equations in \nref{eq: real nominal average dynamics} is continuous, the system is a well-posed hybrid dynamical system \cite[Thm. 6.30]{goebel2009hybrid} and therefore the $O(\bar a)$ perturbed system in \nref{eq: average dynamics with O(a)} renders the set $\mathcal{A} \times \mathcal{K}^q \times \mathcal{S}^q \times \R^m$ SGPAS as $(\bar \varepsilon, \bar a)\rightarrow 0$ \cite[Prop. A.1]{poveda2021robust}. By noticing that the set $\mathbb{S}^{m}$ is UGAS under oscillator dynamics in \nref{eq: oscilator} that generate a well-defined average system in \nref{eq: average dynamics with O(a)}, and by averaging results in \cite[Thm. 7]{poveda2020fixed}, we obtain that the dynamics in \nref{eq: zeroth-order dynamics} make the set $\mathcal{A} \times \mathcal{K}^q \times \mathcal{S}^q \times \R^m \times \mathbb{S}^{m}$ SGPAS as $(\bar \varepsilon, \bar a, \bar \nu) \rightarrow 0$ for the restricted flow and jump sets. Furthermore, by \cite[Prop. A.1.]{poveda2021robust}, HDS $((C \cap \mathscr{K}) \times \R^m \times \mathbb{S}^m), (D \cap \mathscr{K}) \times \R^m \times \mathbb{S}^m, F_0, G_0)$ is structurally robust.

\end{document}

%% file: comande.tex
\usepackage{graphicx}
\usepackage{amsmath}
\usepackage{amssymb}
\usepackage{mathtools} 
\usepackage[utf8]{inputenc}
\usepackage{letltxmacro}
\usepackage[table,xcdraw]{xcolor}
\usepackage{hhline} 

\newcommand{\nref}[1]{(\ref{#1})}
\newcommand{\bfs}[1]{\boldsymbol{#1}}
\newcommand{\col}[1]{\operatorname{col}\left({#1}\right)}
\newcommand{\diag}[1]{\operatorname{diag}\left({#1}\right)}
\newcommand{\kraj}{\hspace*{\fill} \qed}
\newcommand{\krajdokaz}{\hfill $\blacksquare$}

\newcommand{\z}[1]{\left( #1 \right)}
\newcommand{\m}[1]{\begin{bmatrix} #1 \end{bmatrix}}
\newcommand{\n}[1]{\left\| #1 \right\|}
\newcommand{\red}{\nonumber \\}
\newcommand{\vprod}[2]{\left \langle #1 \ \middle\vert\  #2 \right \rangle}
\newcommand{\cal}[1]{\mathcal{#1}}

\newcommand{\R}{\mathbb{R}}

\newcommand{\dom}{\mathrm{dom}}

\newcommand{\argmin}{\mathrm{argmin}}